\documentclass{emulateapj}

\usepackage{color}
\usepackage{multirow}

\usepackage{natbib}

\shorttitle{Environment and the Evolution of Galaxy Sizes}
\shortauthors{Cooper et al.}

\begin{document}


\title{The DEEP3 Galaxy Redshift Survey: The Impact of Environment on 
  the Size Evolution of Massive Early-type Galaxies at Intermediate
  Redshift\footnotemark[\dag]\footnotemark[\ddag]}


\author{
Michael C.\ Cooper\altaffilmark{1,$\nabla$}
Roger L.\ Griffith\altaffilmark{2},
Jeffrey A.\ Newman\altaffilmark{3},
Alison L.\ Coil\altaffilmark{4,$\exists$},
Marc Davis\altaffilmark{5,6},
Aaron A.\ Dutton\altaffilmark{7,$\flat$},
S.\ M.\ Faber\altaffilmark{8},
Puragra Guhathakurta\altaffilmark{8},
David C.\ Koo\altaffilmark{8},
Jennifer M.\ Lotz\altaffilmark{9},
Benjamin J.\ Weiner\altaffilmark{10},
Christopher N.\ A.\ Willmer\altaffilmark{10},
Renbin Yan\altaffilmark{11}
}

\footnotetext[\dag]{Some of the data presented herein
  were obtained at the W.M.\ Keck Observatory, which is operated as a
  scientific partnership among the California Institute of Technology,
  the University of California and the National Aeronautics and Space
  Administration. The Observatory was made possible by the generous
  financial support of the W.M.\ Keck Foundation.}

\footnotetext[\ddag]{Based on observations made with the NASA/ESA Hubble
  Space Telescope, obtained from the data archive at the Space
  Telescope Science Institute. STScI is operated by the Association of
  Universities for Research in Astronomy, Inc.\ under NASA contract NAS
  5-26555.}

\altaffiltext{1}{Center for Galaxy Evolution, Department of Physics
  and Astronomy, University of California, Irvine, 4129 Frederick
  Reines Hall, Irvine, CA 92697, USA; m.cooper@uci.edu}

\altaffiltext{$\nabla$}{Hubble Fellow}

\altaffiltext{2}{Infrared Processing and Analysis Center, California
  Institute of Technology, Pasadena, CA 91125, USA}

\altaffiltext{3}{Department of Physics and Astronomy, University of
  Pittsburgh, 3941 O'Hara Street, Pittsburgh, PA 15260, USA}

\altaffiltext{4}{Department of Physics, Center for Astrophysics and
  Space Sciences, University of California, San Diego, 9500 Gilman
  Drive, La Jolla, CA, 92093}

\altaffiltext{$\exists$}{Alfred P.\ Sloan Foundation Fellow}

\altaffiltext{5}{Department of Astronomy, 
  University of California, Berkeley, Hearst Field Annex B, Berkeley,
  CA 94720, USA}

\altaffiltext{6}{Department of Physics, University of California,
  Berkeley, 366 LeConte Hall MC 7300, Berkeley, CA 94720, USA}

\altaffiltext{7}{Department of Physics and Astronomy, University of
  Victoria, Victoria B.C., V8P 5C2, Canada}

\altaffiltext{$\flat$}{CITA National Fellow}

\altaffiltext{8}{UCO/Lick Observatory and Department of Astronomy and
  Astrophysics, University of California, Santa Cruz, 1156 High
  Street, Santa Cruz, CA 95064, USA}

\altaffiltext{9}{Space Telescope Science Institute, 3700 San Martin
  Drive, Baltimore, MD 21218, USA}

\altaffiltext{10}{Steward Observatory, University of Arizona, 933 N.\
  Cherry Avenue, Tucson, AZ 85721, USA}

\altaffiltext{11}{Center for Cosmology and Particle Physics,
  Department of Physics, New York University, 4 Washington Place, New
  York, NY 10003, USA}

\begin{abstract}
  Using data drawn from the DEEP2 and DEEP3 Galaxy Redshift Surveys,
  we investigate the relationship between the environment and the
  structure of galaxies residing on the red sequence at intermediate
  redshift. Within the massive $(10 < \log_{10}({\rm M}_{\star} /
  h^{-2}\ {\rm M}_{\sun}) < 11)$ early-type population at $0.4 < z <
  1.2$, we find a significant correlation between local galaxy
  overdensity (or environment) and galaxy size, such that early-type
  systems in higher-density regions tend to have larger effective
  radii (by $\sim \! 0.5\ h^{-1}$ kpc or $25\%$ larger) than their
  counterparts of equal stellar mass and S\'{e}rsic index in
  lower-density environments. This observed size-density relation is
  consistent with a model of galaxy formation in which the evolution
  of early-type systems at $z < 2$ is accelerated in high-density
  environments such as groups and clusters and in which dry, minor
  mergers (versus mechanisms such as quasar feedback) play a central
  role in the structural evolution of the massive, early-type galaxy
  population.

\end{abstract}


\section{Introduction}
\label{sec_intro}

Recent observations of galaxies at intermediate redshift ($z \sim 2$)
have identified a significant population of massive ($\sim \!
10^{11}$ M$_{\sun}$), quiescent, early-type systems with old,
metal-rich stellar populations and remarkably small sizes relative to
their local counterparts \citep[e.g.,][]{daddi05, labbe05, papovich06,
  kriek06, trujillo06, trujillo07, zirm07, vandokkum08, damjanov09,
  damjanov11, toft09, taylor10}. The stellar densities of these
massive, intermediate-redshift galaxies (as measured within one
effective radius, $r_{e}$) are typically two orders of magnitude
greater than quiescent ellipticals of the same mass at $z \sim
0.1$. Within the central $1$ kpc (physical), however, the densities of
early-types at $z \sim 2$ are found to only exceed local measurements
by a factor of $2$--$3$ \citep{bezanson09, hopkins09b,
  vandokkum10}. Altogether, the observations suggest that there is
significant evolution in the size of massive ellipticals over the past
$10$ Gyr, likely proceeding in an inside-out manner, without the
addition of much stellar mass.

Several physical processes have been proposed to explain this strong
size evolution within the massive, early-type population at $z < 2$.
In particular, gas-poor, collisionless (or ``dry'') minor mergers are
often invoked as a means for puffing up the stellar component of these
massive systems \citep[e.g.,][]{naab06, naab07, naab09, khochfar06,
  bournaud07, bk07, vdw09, cenarro09, hopkins09a, hopkins10b,
  trujillo11}. However, a variety of alternative mechanisms have also
been proposed, including scenarios in which the observed structural
evolution may be driven by secular processes such as adiabatic
expansion resulting from stellar mass loss and/or strong AGN-fueled
feedback (\citealt{fan08, fan10}; \citealt{damjanov09};
\citealt{hopkins10a, hopkins10b}; see also \citealt{nipoti09} and
\citealt{williams10}).

One way to possibly discriminate between these scenarios (minor
mergers versus secular processes) is by quantifying the role of
environment in the structural evolution of the massive galaxy
population. While secular processes are largely independent of
environment and quasars are not preferentially found in overdense
regions at $z \sim 1$ \citep{coil07}, mergers are more common in
higher-density environments such as galaxy groups \citep{cavaliere92,
  mcintosh08, wetzl08, fakhouri09, lin10, darg10}. Thus, if mergers
are the dominant mechanism by which the sizes of massive early-types
evolve at $z < 2$, we should expect to find a variation in the
structural properties of galaxies as a function of environment at $z
\sim 1$.

To test this, we use data drawn from the DEEP2 and DEEP3 Galaxy
Redshift Surveys \citep{davis03, newman12, cooper11} to investigate
the correlation between the local overdensity of galaxies (which we
generally refer to as ``environment'') and the sizes of massive
galaxies on the red sequence at intermediate redshift. In Section
\ref{sec_data}, we describe our data set, with results and discussion
presented in Sections \ref{sec_results} and \ref{sec_disc},
respectively. Throughout, we employ a $\Lambda$CDM cosmology with $w =
-1$, $\Omega_m = 0.3$, $\Omega_{\Lambda} = 0.7$, and a Hubble
parameter of $H_{0} = 100\ h$ km s$^{-1}$ Mpc$^{-1}$, unless otherwise
noted. All magnitudes are on the AB system \citep{oke83}.

\section{Data}
\label{sec_data}

To characterize both the environment and the structure of galaxies
accurately requires spectroscopic observations \citep[or deep,
multi-band photometric observations,][]{cooper05} as well as
high-resolution imaging across a sizable area of sky. Given the
limitations of ground-based adaptive-optics observations, the latter
is only possible at intermediate redshift via space-based observations
(e.g., with {\it HST}). Among the fields covered by deep, multi-band
imaging with {\it HST}, the Extended Groth Strip (EGS) is by far the
most complete with regard to spectroscopic coverage at intermediate
redshift. The EGS is one of four fields surveyed by the DEEP2 Galaxy
Redshift Survey \citep{davis03, davis07, newman12}, yielding
high-precision ($\sigma_{z} \sim 30$ km s$^{-1}$) secure redshifts for
$11,701$ sources at $0.2 < z < 1.4$ over roughly $0.5$ square degrees
in the EGS. Building upon the DEEP2 spectroscopic sample, the
recently-completed DEEP3 Galaxy Redshift Survey (\citealt{cooper11};
Cooper et al., in prep) has brought the target sampling rate to $\sim
\! 90\%$ at $R_{\rm AB} < 24.1$ over the central $0.25$ square degrees
of the EGS --- the portion of the field imaged by {\it HST}/ACS
\citep[see Fig.\ \ref{fig_egs};][]{davis07, lotz08}.

Among the current generation of deep spectroscopic redshift surveys at
$z \sim 1$, the combination of the DEEP2 and DEEP3 spectroscopic
datasets provides the largest sample of accurate spectroscopic
redshifts, the highest-precision velocity information, and the highest
sampling density (\citealt{newman12}; Cooper et al., in
prep).\footnote{Note that the sampling density for a survey is defined
  to be the number of galaxies with an accurate redshift measurement
  per unit of comoving volume and \emph{not} the number of galaxies
  targeted down to an arbitrary magnitude limit.}  Combined with the
relatively wide area imaged with {\it HST}/ACS \citep[an area $> \!
2\times$ larger than that surveyed as part of the GOODS
program,][]{giavalisco04}, these attributes make the EGS one of the
best-suited fields in which to study the relationship between
environment and galaxy structure at $z \sim 1$. In this paper, we
utilize a parent sample of $11,493$ galaxies drawn from the joint
DEEP2/DEEP3 dataset in the EGS with secure redshifts \citep[quality $Q
= 3$ or 4 as defined by][]{davis07, newman12} in the range $0.4 < z <
1.2$.

\subsection{Rest-frame Colors, Luminosities, and Stellar Masses}

For each galaxy in the DEEP2/DEEP3 sample, rest-frame $U-B$ colors and
absolute $B$-band magnitudes, $M_{B}$, are calculated from CFHT $BRI$
photometry \citep{coil04b} using the $K$-correction procedure
described in \citet{willmer06}. For a subset of the galaxy catalog,
stellar masses are calculated by fitting spectral energy distributions
(SEDs) to WIRC/Palomar $J$- and $K_{s}$-band photometry in conjunction
with the DEEP2 $BRI$ data, according to the prescriptions described by
\cite{bundy05, bundy06}. However, the near-infrared photometry,
collected as part of the Palomar Observatory Wide-field Infrared
\citep[POWIR,][]{conselice08} survey, does not cover the entire
DEEP2/DEEP3 survey area, and often faint blue galaxies at the high-$z$
end of the DEEP2 redshift range are not detected in $K_{s}$. Because
of these two effects, the stellar masses of \citet{bundy06} have been
used to calibrate stellar mass estimates for the full DEEP2 sample
that are based on rest-frame $M_{B}$ and $B-V$ values derived from the
DEEP2 data in conjunction with the expressions of \citet{bell03},
which relate mass-to-light ratio to optical color. We empirically
correct these stellar mass estimates to the \citet{bundy06}
measurements by accounting for a mild color and redshift dependence
\citep{lin07}; where they overlap, the two stellar masses have an rms
difference of approximately $0.3$ dex after this calibration.

\subsection{Local Galaxy Overdensity}

To characterize the local environment, we compute the projected
third-nearest-neighbor surface density $(\Sigma_3)$ about each galaxy
in the joint DEEP2/DEEP3 sample, where the surface density depends on
the projected distance to the third-nearest neighbor, $D_{p,3}$, as
$\Sigma_3 = 3 / (\pi D_{p,3}^2)$. In computing $\Sigma_3$, a velocity
window of $\pm 1250$ km s$^{-1}$ is utilized to exclude foreground and
background galaxies along the line of sight. Varying the width of this
velocity window (e.g., using $\pm 1000$--$2000$ km s$^{-1}$) or
tracing environment according to the projected distance to the
fifth-nearest neighbor has no significant effect on our results. In
the tests of \citet{cooper05}, this projected $n^{\rm
  th}$-nearest-neighbor environment estimator proved to be the most
robust indicator of local galaxy density for the DEEP2 survey.

\begin{figure}[h!]
\centering
\plotone{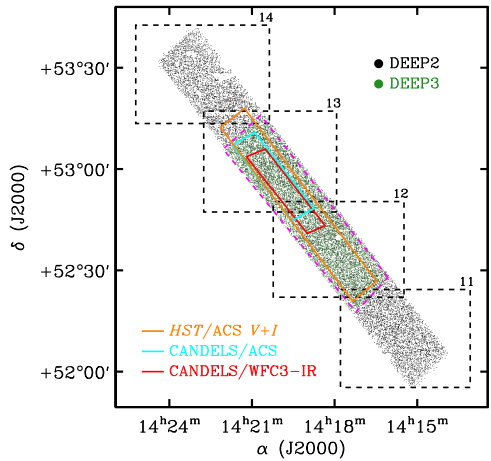}
\caption{DEEP2 and DEEP3 spectroscopic coverage in the Extended Groth
  Strip. The black and green points denote the spectroscopic targets
  of the two surveys, with the area covered by {\it HST}/ACS imaging
  highlighted in orange. The dashed magenta line denotes the edge of
  the DEEP3 survey and thus the area over which environments were
  computed for this work. The areas to be imaged with {\it HST}/ACS
  and WFC3-IR as part of the CANDELS Multi-Cycle Treasury Program
  \citep{grogin11} are delineated by the cyan and red outlines,
  respectively. Finally, the location of the four CFHT pointings
  (numbered $11$--$14$) imaged by \citet{coil04b} are denoted by the
  black dashed lines.}
\label{fig_egs}
\end{figure}

To correct for the redshift dependence of the DEEP2 and DEEP3 sampling
rates, each surface density value is divided by the median $\Sigma_3$
of galaxies at that redshift within a window of $\Delta z = 0.04$;
correcting the measured surface densities in this manner converts the
$\Sigma_3$ values into measures of overdensity relative to the median
density (given by the notation $1 + \delta_{3}$ here) and effectively
accounts for the redshift variations in the selection rate
\citep{cooper05}. Finally, to minimize the effects of edges and holes
in the survey geometry, we exclude all galaxies within $1$ $h^{-1}$
comoving Mpc of the DEEP3 survey boundary (see Fig.\ \ref{fig_egs}),
reducing our sample to $7,257$ galaxies in the redshift range $0.4 < z
< 1.2$.

\begin{figure}[h!]
\centering
\plotone{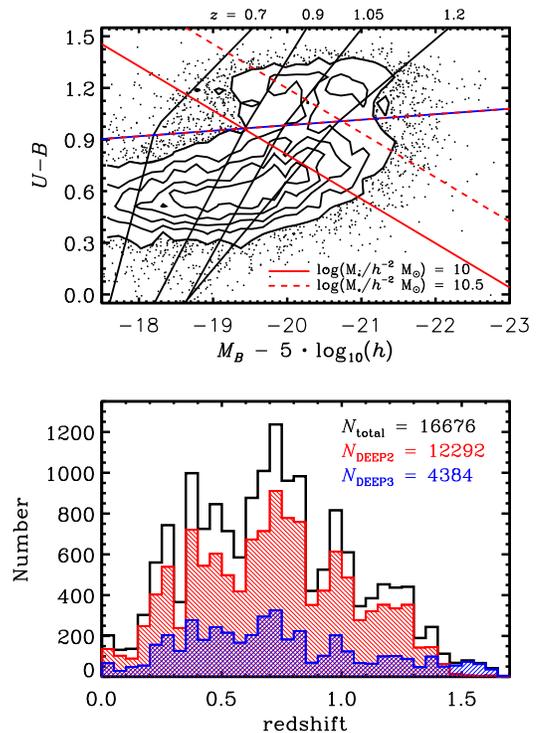}
\caption{\emph{Top}: the rest-frame $U-B$ versus $M_B$ color-magnitude
  distribution for the DEEP2 and DEEP3 galaxies in the spectroscopic
  sample within the redshift range $0.4 < z < 1.2$. The four solid
  black lines show the effective absolute-magnitude limit of the
  survey at $z = 0.7$, $0.9$, $1.05$, and $1.2$, while the solid and
  dashed red lines show lines of constant stellar mass corresponding
  to $\log_{10}({\rm M}_{\star}/h^{-2}\ {\rm M}_{\sun}) = 10$ and
  $10.5$, respectively. The dashed blue/red horizontal line shows the
  division between the red sequence and the blue cloud as given by
  Equation 19 of \citet{willmer06}. \emph{Bottom}: the combined
  distribution of the $16,676$ unique, secure ($Q = -1$, $3$, $4$)
  redshifts measured by DEEP2 and DEEP3 within the Extended Groth
  Strip field (black histogram). The red and blue histograms show the
  corresponding distributions for the DEEP2 and DEEP3 surveys
  independently. Recall that the DEEP2 survey covers roughly a factor
  of $2$ more area than the DEEP3 survey (see Fig.\ \ref{fig_egs}). }
\label{fig_cmdz}
\end{figure}

\subsection{S\'{e}rsic Indices and Sizes}

To quantify the sizes of the DEEP2 and DEEP3 galaxies, we utilize
morphological measurements extracted from the Advanced Camera for
Surveys General Catalog (ACS-GC, Griffith et al.\ 2012, in prep). The
ACS-GC analyzed the {\it HST}/ACS $V_{\rm F606W}$ and $I_{\rm F814W}$
imaging in the EGS using GALAPAGOS \citep{haussler11}, a
semi-automated tool for measuring sizes and spatial profiles via the
parametric fitting code GALFIT \citep{peng02, peng10}. To determine
the galaxy profile shape, each radial profile was fit using a simple
S\'{e}rsic measurement \citep{sersic68} of the form
\begin{equation}
\Sigma(r) = \Sigma_{e} \: {\rm exp}[-\kappa ((r/r_{e})^{-n} -1)]
\end{equation}
where $r_{e}$ is the effective radius of the galaxy, $\Sigma_{e}$ is
the surface brightness at $r_{e}$, $n$ is the power-law index, and $\kappa$
is coupled to $n$ such that half of the total flux is always within
$r_e$. 

Here, we employ the profile fits to the {\it HST}/ACS $I_{\rm F814W}$
imaging for all sources, independent of galaxy redshift. At $z < 1.2$,
the $I_{\rm F814W}$ passband samples the rest-frame optical ($\lambda
> 3700$\AA), which minizes morphological biases associated with
observations made in the rest-frame ultraviolet where galaxies
typically exhibit more irregular morphologies. The impact (or lack
thereof) of any morphological $K$-correction is addressed further in
\S \ref{sec_results}. Throughout our analysis, all sizes ($r_{e}$)
have been converted to physical kpc, according to the DEEP2/DEEP3
spectroscopic redshift and assuming a Hubble parameter of $h =
1$. Finally, note that the multidrizzled {\it HST}/ACS images, from
which structural properties were measured, have a pixel scale of
$0.03^{\prime\prime}$ per pixel and a point-spread function (PSF) of
$0.12^{\prime\prime}$ FWHM; from $z=0.4$ to $z = 1.2$, the spatial
resolution therefore varies from $\sim \! 0.5$ $h^{-1}$ kpc per PSF FWHM
to $\sim \!  0.95$ $h^{-1}$ kpc per PSF FWHM.

\subsection{Sample Selection}

To investigate the relationship between galaxy structure and
environment amongst the high-mass portion of the red sequence, we
define a subsample of DEEP2/DEEP3 galaxies at $0.4 < z < 1.2$ with
stellar mass in the range $10 < \log_{10}({\rm M}_{\star} / h^{-2}\
{\rm M}_{\sun}) < 11$, on the red sequence (i.e., rest-frame color of
$U -B > 1$),\footnote{We adopt this simplified color-cut to be more
  restrictive at the faint end of the red sequence, where dusty
  star-forming galaxies are more prominent \citep{lotz08}. However,
  using a luminosity-dependent color-cut \citep[e.g., Equation 19
  of][]{willmer06} yields no significant changes in our results.} and
with robust environment and morphology measurements (i.e., away from a
survey edge and with $\sigma_{n} < 0.75$).\footnote{Limiting the
  sample to those sources with $\sigma_{n} < 0.75$ excludes very few
  (only $2$ out of $625$) objects. Removing this selection criterion
  or making it more restrictive (e.g., $\sigma_{n} < 0.5$) yields no
  significant changes in our results.} The median redshift for this
subsample of $623$ galaxies is $0.76$ and the median stellar mass is
$\log_{10}({\rm M}_{\star} / h^{-2}\ {\rm M}_{\sun}) \sim 10.6$.

In Figure \ref{fig_cmdz}, we show the redshift distribution for all
sources in the EGS with a secure redshift ($Q = -1$, $3$, $4$) in the
joint DEEP2/DEEP3 sample alongside the color-magnitude distribution
for all galaxies at $0.4 < z < 1.2$, with lines of constant stellar
mass overlaid and with lines illustrating the survey magnitude limit
at several discrete redshift values. Note that we restrict our primary
subsample to a redshift range over which the DEEP2 and DEEP3 selection
function is relatively flat. However, over this somewhat broad
redshift range the sample is incomplete at the adopted mass limit. For
example, at $z=0.9$ the $R_{\rm AB}=24.1$ magnitude limit of DEEP2
includes all galaxies with stellar mass $> \! 10^{10.8}\ {\rm
  M}_{\star}/h^{-2}\ {\rm M}_{\sun}$ independent of color, but
preferentially misses red galaxies at lower masses (see Fig.\
\ref{fig_cmdz}). This incompleteness in the galaxy population is
addressed in more detail in \S \ref{sec_results}.

\section{Analysis}
\label{sec_results}

In order to study the relationship between galaxy properties and
environment at fixed stellar mass, as we aim to do here, the galaxy
sample under study is often restricted to a narrow range in stellar
mass such that correlations between stellar mass and environment are
negligible. However, at intermediate redshift, sample sizes are
generally limited in number such that using a particularly narrow
stellar mass range (e.g., $\sim \!  0.1$--$0.2$ dex in width)
significantly reduces the statistical power of the sample. For this
reason, broader stellar mass bins (e.g., $\sim \! 0.5$ dex) are
commonly employed. However, if the shape of the stellar mass function
depends on environment (as suggested by \citealt{balogh01,
  kauffmann04, croton05, baldry06, rudnick09, cooper10b,
  bolzonella10}), then the typical stellar mass within a broad mass
bin may differ significantly from one density regime to another. Such
an effect would clearly impact the ability to study the relationship
between galaxy size and environment at fixed stellar mass.

For this reason, we instead select those galaxies within the top
$15\%$ of the overdensity distribution for all red galaxies at $10 <
\log_{10}({\rm M}_{\star} / h^{-2}\ {\rm M}_{\sun}) < 11$ and $0.4 < z
< 1.2$ (a subsample of $93$ galaxies), and from the corresponding
bottom $50\%$ of the overdensity distribution we randomly draw 1000
subsamples (each composed of $93$ galaxies) so as to match the joint
redshift, stellar mass, and S\'{e}rsic index distributions of the
galaxies in the high-density subsample.\footnote{Selecting the top
  10\% or 20\% of the environment distribution yields similar
  results.}  The average environment for the high-density subset is
$\log_{10}(1+\delta_3) = 1.31$, with an interquartile (25\%--75\%) range of
$1.15$--$1.41$, while the low-density subsample is biased to
considerably less-dense environs with an average overdensity of
$\log_{10}(1+\delta_3) = -0.12$ and an interquartile range of
$-0.30$--$0.12$. As shown by \citet{cooper06}, the high-density
subsample is comprised largely of group members, while the low-density
population is dominated by ``field'' galaxies \citep[see
also][]{cooper07, gerke07}.

By matching in redshift, we remove the projection of any
possible residual correlation between our environment measurements and
redshift (in concert with the known redshift dependence of the
survey's stellar-mass limit) onto the observed size-environment
relation. In addition, matching according to redshift alleviates any
possible impact from morphological $K$-corrections associated with
measuring all structural parameters in the {\it HST}/ACS $I_{\rm
  F814W}$ passband. Finally, recognizing the correlations between
environment and parameters such as color, star-formation rate, and
morphology (i.e., early- versus late-type) at $z \sim 1$
\citep[e.g.,][]{cooper06, cooper08, cooper10b, capak07, elbaz07,
  vdw07}, we also force the S\'{e}rsic indices of the low-density
subsample to match those of the high-density population, which
controls for the contribution of dusty disk galaxies to our
red-sequence population. Matching based on rest-frame color, in lieu
of S\'{e}rsic index, would confuse reddened disk galaxies with red
early-type systems.

Members of the low-density subsample are drawn randomly from within a
three-dimensional window with dimensions of $|\Delta z| < 0.04$,
$|\Delta \log_{10}({\rm M}_{\star} / h^{-2}\ {\rm M}_{\sun})| < 0.2$,
and $|\Delta n| < 1.5$ of a randomly-selected object in the
high-density subsample. Varying the size of this window by factors of
a few in each dimension has no significant effect on our
results. Given the random nature of the matching, some objects are
repeated in the low-density subsamples. However, for each subsample of
$93$ galaxies, $> \! 70\%$ of the galaxies are unique; requiring all
members of a subsample to be unique would skew the statistics
\citep{efron81}. By matching our high- and low-density subsamples in
stellar mass, redshift, as well as S\'{e}rsic index, we are able to
effectively study the correlation between galaxy size and environment
at fixed stellar mass.

To test whether our high-density subsample and the random low-density
subsamples are consistent with being drawn from the same underlying
stellar mass distribution, we apply two non-parametric (i.e.,
independent of Gaussian assumptions) tests, the two-sided
Kolmogorov-Smirnov (KS) test \citep{press86, wall03} and the one-sided
Wilcoxon-Mann-Whitney (WMW) $U$ test \citep{mann47}. The result of
each test is a $P$-value: the probability that a value of the KS or
$U$ statistic equal to the observed value or more extreme would be
obtained, if the ``null'' hypothesis holds that the samples are drawn
from the same parent distribution. The WMW $U$ test is computed by
ranking all elements of the two data sets together and then comparing
the mean (or total) of the ranks from each data set. Because it relies
on ranks rather than observed values, it is highly robust to
non-Gaussianity. The WMW $U$ test is particularly useful for small
data sets (e.g., compared to other related tests such as the
chi-square two-sample test, \citealt{wall03}), as we have when
selecting galaxies from a narrow stellar mass range and in extreme
environments, due to its insensitivity to outlying data points, its
avoidance of binning, and its high efficiency. Note that since this
test is one-sided, possible $P_{U}$ values range from $0$ to $0.5$
(versus $P_{\rm KS}$ which ranges from $0$ to $1$); for a $P_{U}$
value below $0.025$ (corresponding closely to $2\sigma$ for a
Gaussian), we can reject the null hypothesis (that the two samples
have the same distribution) at greater than $95\%$
significance.\footnote{Note that the two-sided probability for the WMW
  $U$ test is given by doubling the one-sided probability. Here, we
  report only the one-sided probability.}

\begin{figure}[h]
\centering
\plotone{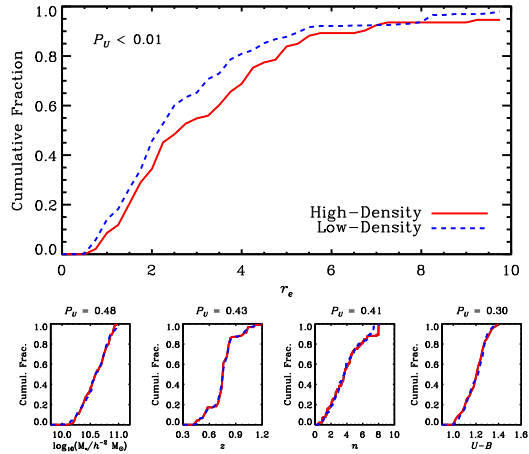}
\caption{the cumulative size ($r_{e}$), stellar mass, redshift,
  S\'{e}rsic index ($n$), and rest-frame color ($U-B$) distributions
  for the $93$ DEEP2 galaxies comprising the top $15\%$ of the
  environment distribution within the stellar mass and redshift ranges
  of $10 < \log_{10}({\rm M}_{*} / h^{-2}\ {\rm M}_{\sun}) < 11$ and
  $0.4 < z < 1.2$ in comparison to the corresponding cumulative
  distributions for the $1000$ random galaxy subsamples drawn from the
  lowest $50\%$ of the same environment distribution. As discussed in
  the text, the low-density subsamples, which are each composed of
  $93$ galaxies, are selected to have the same stellar mass,
  redshift, and S\'{e}rsic index distribution as the high-density
  population. However, the size distribution is found to be
  significantly different in the different environment regimes.}
\label{fig_cdists}
\end{figure}

In Figure \ref{fig_cdists}, we plot the cumulative distribution of
stellar masses for the $93$ sources in the high-density subsample
alongside that for the 1000 random subsamples (each consisting of $93$
galaxies) matched in redshift but residing in low-density
environments. Performing a one-sided WMW $U$ test (and a two-sided KS
test) on the size ($r_{e}$) measurements for the low- and high-density
populations, we find that the size distribution for the galaxies in
high-density environments is skewed to larger sizes, with a
probability of $P_{U} < 0.01$ (and $P_{\rm KS} < 0.02$). Meanwhile,
the cumulative stellar mass, redshift, and S\'{e}rsic index
distributions for the low- and high-density subsamples, shown in the
inset of Figure \ref{fig_cdists}, are well-matched with the WMW $U$
test yielding a $P_{U} > 0.4$. This confirms that our
sample-construction procedure has yielded sets of galaxies in low- and
high-density environments whose redshift, mass, and S\'{e}rsic index
distributions match closely. While not directly matched, the
rest-frame color distributions for the two samples are also
indistinguishable --- not a surprising result given that the
color-density relation shows no significant variation across the red
sequence at a given luminosity \citep{blanton05, cooper06}. See Table
\ref{res_tab1} for a complete summary of the probability values given
by both the WMW $U$ and KS tests.

The results of the WMW $U$ test are confirmed by a comparison of the
Hodges-Lehmann (H-L) estimator of the mean sizes for the low- and
high-density subsamples, which differ by $0.54 \pm 0.22$ $h^{-1}$
kpc. This reinforces the conclusion that there is a non-negligible
size-environment relation on the red sequence at $z \sim 0.75$. The
Hodges-Lehmann (H-L) estimator of the mean is given by the median
value of the mean computed over all pairs of galaxies in the sample
\citep{hodges63}. Like taking the median of a distribution, the H-L
estimator of the mean is robust to outliers, but, unlike the median,
yields results with scatter (in the Gaussian case) comparable to the
arithmetic mean. Thus, by using the H-L estimator of the mean, we gain
robustness as in the case of the median, but unlike the median, our
measurement errors are increased by only a few percent.

\begin{figure}[h]
\centering
\plotone{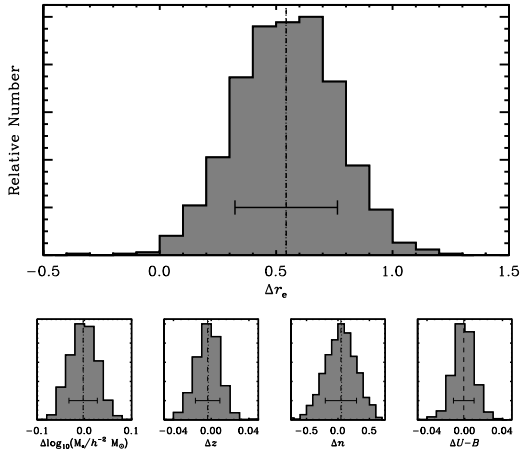}
\caption{the distribution of differences between the Hodges-Lehmann
  (H-L) estimator of the mean size, stellar mass, redshift, S\'{e}rsic
  index, and rest-frame color for the high-density subsample of $93$
  red-sequence galaxies at $0.4 < z < 1.2$ and $10 < \log_{10}({\rm
    M}_{\star} / h^{-2}\ {\rm M}_{\sun}) < 11$ relative to the
  corresponding H-L estimator of the mean for each of the $1000$
  low-density subsamples. The dotted and dashed vertical lines show
  the median and mean, respectively, of the distribution of
  differences between the H-L means for each galaxy property; error
  bars denote the uncertainty in the mean as given in Table
  \ref{res_tab2}. We find a significant offset in size of $\Delta
  r_{e} \sim 0.6\ h^{-1}$ kpc (physical), while the difference in mean
  stellar mass, redshift, S\'{e}rsic index, and color of the two
  samples is consistent with zero (by construction for all but
  color).}
\label{fig_hlmeans}
\end{figure}

In Figure \ref{fig_hlmeans}, we show the distribution of the
differences between the Hodges-Lehmann estimator of the mean size,
stellar mass, redshift, S\'{e}rsic index, and color for the
high-density subsample relative to that for each of the $1000$
low-density subsamples, where the median difference in the H-L
estimate of the mean size is $\Delta r_{e} = 0.559$ (as illustrated by
the dashed vertical line) versus $\Delta \log_{10}({\rm M}_{\star} /
h^{-2}\ {\rm M}_{\sun}) = -0.002$, $\Delta z = -0.003$, $\Delta n =
0.057$, and $\Delta (U-B) = -0.001$ for stellar mass, redshift,
S\'{e}rsic index, and color, respectively (see Table
\ref{res_tab2}). Within the stellar mass range of $10 < \log_{10}({\rm
  M}_{\star} / h^{-2}\ {\rm M}_{\sun}) < 11$, we find significant
evidence for a correlation between galaxy size and environment at $z
\sim 0.75$, such that higher-density regions play host to larger
galaxies at a given stellar mass on the red sequence.

To test the robustness of our results to the particularities of the
sample selection, we repeat the analysis described above for several
samples spanning varying redshift and stellar mass regimes. For
example, restricting the redshift range over which we select galaxies
to $0.7 < z < 1.2$, thereby decreasing the size of the sample, we
again find a statistically significant relationship between galaxy
structure and environment within our adopted stellar mass bin of $10 <
\log_{10}({\rm M}_{\star} / h^{-2}\ {\rm M}_{\sun}) < 11$. For the
$64$ galaxies in the high-density regime (again the highest $15\%$ of
the overdensity distribution) at $0.7 < z < 1.2$, the cumulative
distribution of galaxy size ($r_{e}$) is skewed towards larger
effective radii relative to the comparison set of galaxies in
low-density environments, yielding $P_{U} < 0.01$ and a median
difference in the H-L estimator of the mean size of $\Delta r_{e} \sim
0.63$ $h^{-1}$ kpc.

An obvious concern when studying the structure of galaxies on the red
sequence is the relative contribution of early-type systems and dusty
disk galaxies to the sample. Especially at fainter magnitudes/lower
stellar mass, reddened star-forming galaxies comprise a significant
portion of the red galaxy population \citep[e.g.,][]{lotz08, bundy10,
  cheng11}, and as many studies of environment and clustering at
intermediate redshift have shown the star-forming population tends to
reside in lower-density regions relative to their passive counterparts
\citep[e.g.,][]{cooper06, cooper07, capak07, coil08, kovac10}. Even
with our primary sample selected to be at the massive end of the red
sequence, disk galaxies ($n < 2.5$) still account for $\sim \! 25\%$
of the population (see Figure \ref{fig_cdists}). It is unlikely,
however, that a difference in the relative contribution of dusty disk
galaxies to the high- and low-density populations is driving the
observed size-density relation since galaxy size correlates with
S\'{e}rsic index such that red disk galaxies (systems with $n < 2.5$)
tend to have slightly larger (not smaller) measured sizes than red
galaxies of slightly larger S\'{e}rsic index. Still, to minimize any
potential impact from late-type systems, we define two subsamples: one
selected according to $n > 2.5$ and a second further constrained to
systems with $n > 2.5$ and with axis ratios, $(b/a)$, greater than
$0.4$ (see Table \ref{res_tab1}). For these subsamples of galaxies
with early-type morphology, we find that the correlation between
structure and environment persists, reinforcing the conclusion that
early-type systems (and not dusty disk galaxies) are responsible for
the observed size-density relation at fixed stellar mass.

\begin{figure}[h]
\centering
\plotone{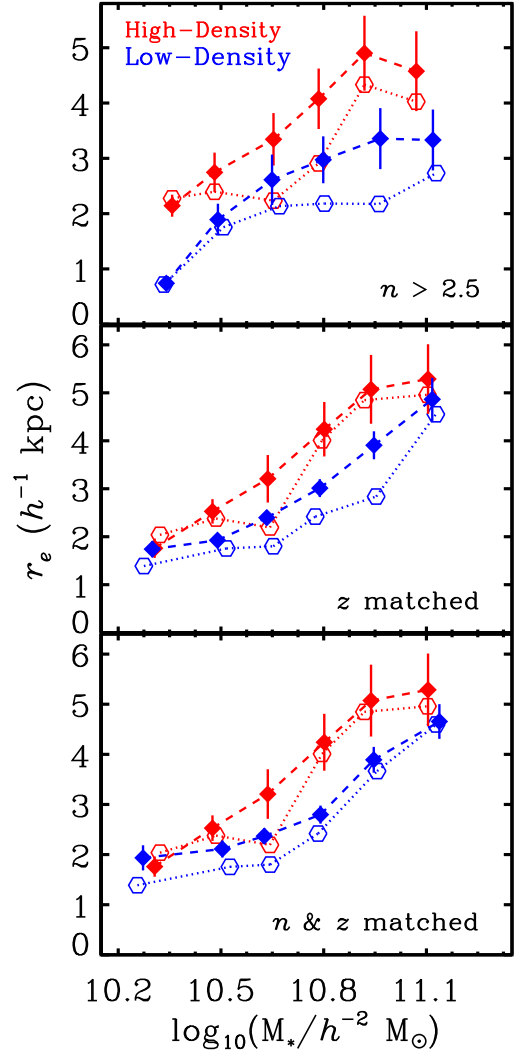}
\caption{the mean (filled diamond connected by dashed lines) and
  median (open hexagons connected by dotted lines) relationships
  between galaxy size ($r_{e}$) and stellar mass in the high-density
  (red lines and symbols) and low-density (blue lines and symbols)
  regimes. The means and medians are computed in bins of stellar mass
  with width $\Delta \log_{10}({\rm M}_{\star} / h^{-2}\ {\rm
    M}_{\sun}) = 0.3$. In the \emph{top} panel, the high- and
  low-density samples are selected as the respective extreme $25\%$ of
  the environment distribution for all galaxies with $U-B > 1$, $0.7 <
  z < 1.2$, and $n > 2.5$ --- with no matching according to $n$ or
  $z$. In the \emph{middle} and \emph{bottom} panels, the high-density
  sample comprises the top $15\%$ of the environment distribution for
  all galaxies with $U-B > 1$, $0.4 < z < 1.2$, and $n > 2.5$, with
  the low-density sample selected from the bottom $50\%$ of the
  environment distribution so as to have the same $z$ (\emph{and }$n$)
  distributions (but not stellar mass). In both instances, we find
  that the size-stellar mass relation for early-type galaxies is
  systematically offset to larger sizes in overdense regions.}
\label{fig_smrel}
\end{figure}

To illustrate the correlation between size and environment in a more
physically intuitive manner, in Figure \ref{fig_smrel}, we show the
size-stellar mass relations for early-type galaxies in overdense and
underdense regions. In the top portion of Fig.\ \ref{fig_smrel}, the
high- and low-density samples are simply defined to be the extreme
quartiles of the environment distribution for all galaxies with $n >
2.5$, $0.7 < z < 1.2$, and $U-B > 1$. In the middle and bottom panel,
however, a more controlled comparison is made, with the high-density
sample selected as the top $15\%$ of the environment distribution for
all galaxies with $n > 2.5$, $0.4 < z < 1.2$, and $U-B > 1$ and the
low-density sample is comprised of $\sim \! 400$ galaxies randomly
drawn from the bottom $50\%$ of the environment distribution to match
the $z$ distribution (middle) and the joint $n$ and $z$ distribution
(bottom) of the high-density sample. In agreement with our previous
analysis, we find that within the high-mass segment of the early-type
galaxy population the size-stellar mass relation varies with
environment, such that galaxies in high-density regions are larger at
a given stellar mass.

Finally, at low surface brightness levels (i.e., low signal-to-noise
per pixel in the {\it HST}/ACS imaging), GALFIT tends to underestimate
both galaxy size ($r_{e}$) and S\'{e}rsic index. While such effects
are minimal for objects brighter than the sky background
\citep{haussler07}, we define a subsample limited to those sources
with surface brightness in the {\it HST}/ACS $I_{\rm F814W}$ passband
of $\mu < 23.5$ magnitudes per arcsec$^{2}$ ($\mu_{\rm sky} \sim
27.4$), which excludes $58$ of $623$ red galaxies at $0.4 < z < 1.2$
and $10 < \log_{10}({\rm M}_{\star} / h^{-2}\ {\rm M}_{\sun}) <
11$. For our samples with closely matched stellar mass, $z$, and color
(i.e., effectively matched apparent magnitude), such a cut on surface
brightness imposes an upper limit on physical size, which therefore
may impact the observed strength of the size-density
relation. However, in spite of this conservative surface brightness
limit, we still find a significant relationship between size and local
overdensity on the red sequence, with a median difference in the H-L
estimator of the mean size of $\Delta r_{e} \sim 0.47$ $h^{-1}$
kpc. In Table \ref{res_tab1} and Table \ref{res_tab2}, we list the
results from similar analyses of several different galaxy
samples. When varying the S\'{e}rsic index, stellar mass, surface
brightness, axis ratio, and/or redshift regimes probed, we continue to
find a significant size-density relation at fixed stellar mass on the
red sequence.

\begin{deluxetable*}{c c  c c  c c  c c  c c  c c}
\tablecolumns{12}
\tablecaption{\label{res_tab1} Summary of KS and WMW $U$ Probabilities} 
\tablehead{ \multirow{2}{*}{Sample} & \multirow{2}{*}{$N_{\rm high-densty}$} & 
\multicolumn{2}{c}{$\log_{10}({\rm M}_{\star})$} &
\multicolumn{2}{c}{$z$} & \multicolumn{2}{c}{$n$} & 
\multicolumn{2}{c}{$U-B$} & \multicolumn{2}{c}{$r_{e}$} \\
 & & $P_{\rm KS}$ & $P_{U}$ & 
$P_{\rm KS}$ & $P_{U}$ & 
$P_{\rm KS}$ & $P_{U}$ & 
$P_{\rm KS}$ & $P_{U}$ &
$P_{\rm KS}$ & $P_{U}$ 
}
\startdata
\hline
$0.4 < z < 1.2$ & \multirow{2}{*}{$93$} & 
\multirow{2}{*}{$0.718$} & \multirow{2}{*}{$0.480$} & 
\multirow{2}{*}{$0.605$} & \multirow{2}{*}{$0.428$} &
\multirow{2}{*}{$0.220$} & \multirow{2}{*}{$0.413$} & 
\multirow{2}{*}{$0.542$} & \multirow{2}{*}{$0.304$} & 
\multirow{2}{*}{$0.011$} & \multirow{2}{*}{$0.006$} \\
$10 < \log_{10}({\rm M}_{\star}) < 11$ &  &  
 &  &  &  &  &  &  &  &  & \\
\hline
$0.7 < z < 1.2$ & \multirow{2}{*}{$64$} & 
\multirow{2}{*}{$0.887$} & \multirow{2}{*}{$0.421$} &
\multirow{2}{*}{$0.578$} & \multirow{2}{*}{$0.356$} & 
\multirow{2}{*}{$0.562$} &  \multirow{2}{*}{$0.461$} &
\multirow{2}{*}{$0.321$} & \multirow{2}{*}{$0.280$} & 
\multirow{2}{*}{$0.036$}  & \multirow{2}{*}{$0.002$} \\
$10 < \log_{10}({\rm M}_{\star}) < 11$ &  &  
  &  &  &  &  &  &  &  &  &  \\
\hline
$0.4 < z < 1.05$ & \multirow{2}{*}{$86$} & 
\multirow{2}{*}{$0.807$} & \multirow{2}{*}{$0.493$} & 
\multirow{2}{*}{$0.740$} & \multirow{2}{*}{$0.445$} &
\multirow{2}{*}{$0.330$} & \multirow{2}{*}{$0.395$} & 
\multirow{2}{*}{$0.614$} & \multirow{2}{*}{$0.387$} & 
\multirow{2}{*}{$0.042$} & \multirow{2}{*}{$0.020$} \\
$10 < \log_{10}({\rm M}_{\star}) < 11$ &  &  
  &  &  &  &  &  &  &  &  &  \\
\hline
$0.4 < z < 1.2$ & \multirow{2}{*}{$58$} & 
\multirow{2}{*}{$0.795$} & \multirow{2}{*}{$0.494$} & 
\multirow{2}{*}{$0.868$} & \multirow{2}{*}{$0.452$} &
\multirow{2}{*}{$0.220$} & \multirow{2}{*}{$0.391$} & 
\multirow{2}{*}{$0.379$} & \multirow{2}{*}{$0.246$} & 
\multirow{2}{*}{$0.015$} & \multirow{2}{*}{$0.013$} \\
$10.5 < \log_{10}({\rm M}_{\star}) < 11$ &  &  
 &  &  &  &  &  &  &  &  &  \\
\hline
$0.7 < z < 1.2$ & \multirow{3}{*}{$42$} & 
\multirow{3}{*}{$0.729$} & \multirow{3}{*}{$0.475$} & 
\multirow{3}{*}{$0.725$} & \multirow{3}{*}{$0.313$} & 
\multirow{3}{*}{$0.230$} & \multirow{3}{*}{$0.384$} & 
\multirow{3}{*}{$0.301$} & \multirow{3}{*}{$0.298$} & 
\multirow{3}{*}{$0.007$} & \multirow{3}{*}{$0.002$} \\
$10 < \log_{10}({\rm M}_{\star}) < 11$ &  &  
 &  &  &  &  &  &  &  &  & \\
$n > 2.5$ &  &  
 &  &  &  &  &  &  &  &  & \\
\hline
$0.7 < z < 1.2$ & \multirow{3}{*}{$36$} & 
\multirow{3}{*}{$0.711$} & \multirow{3}{*}{$0.476$} & 
\multirow{3}{*}{$0.654$} & \multirow{3}{*}{$0.377$} & 
\multirow{3}{*}{$0.252$} & \multirow{3}{*}{$0.327$} & 
\multirow{3}{*}{$0.649$} & \multirow{3}{*}{$0.393$} & 
\multirow{3}{*}{$0.002$} & \multirow{3}{*}{$0.001$} \\
$10 < \log_{10}({\rm M}_{\star}) < 11$ &  &  
 &  &  &  &  &  &  &  &  &  \\
$n > 2.5$, $b/a > 0.4$ &  &  
 &  &  &  &  &  &  &  &  &  \\
\hline
$0.4 < z < 1.2$ & \multirow{3}{*}{$85$} & 
\multirow{3}{*}{$0.810$} & \multirow{3}{*}{$0.499$} & 
\multirow{3}{*}{$0.525$} & \multirow{3}{*}{$0.453$} & 
\multirow{3}{*}{$0.594$} & \multirow{3}{*}{$0.411$} & 
\multirow{3}{*}{$0.796$} & \multirow{3}{*}{$0.364$} & 
\multirow{3}{*}{$0.059$} & \multirow{3}{*}{$0.010$} \\
$10 < \log_{10}({\rm M}_{\star}) < 11$ &  &  
 &  &  &  &  &  &  &  &  &  \\
$\mu < 23.5$ &  &  
 &  &  &  &  &  &  &  &  &  \\
\hline
\enddata
\tablecomments{For several galaxy samples, we list the results of the
  KS and WMW $U$ tests ($P_{\rm KS}$ and $P_{U}$, respectively) from a
  comparison of the stellar mass, redshift, S\'{e}rsic index, color,
  and size distributions of the respective low- and high-density
  samples. The $P$-values indicate the probability that differences in
  the distribution of the stated quantity as large as those observed
  (or larger) would occur by chance if the two samples shared
  identical distributions. Recall that the $P_{U}$ values are
  one-sided probabilities, while the $P_{\rm KS}$ are two-sided. The
  number of galaxies in the high-density sample (picked to be the top
  15\% of the environment distribution) is given by $N_{\rm
    high-density}$. Note that stellar masses and sizes are in units of
  $h^{-2}\ {\rm M}_{\sun}$ and $h^{-1}$ physical kpc,
  respectively. The samples listed here are matched in stellar mass,
  redshift, and S\'{e}rsic index, by construction.}
\end{deluxetable*}

\begin{deluxetable*}{c  cc   cc   cc   cc   cc}
\tablecolumns{11}
\tablecaption{\label{res_tab2} Summary of Differences between
  Hodges-Lehmann Estimates of the Mean} 
\tablehead{ \multirow{2}{*}{Sample} & \multicolumn{5}{c}{$\Delta_{\rm HL}$} &
  \multicolumn{5}{c}{median of $\Delta_{\rm HL}$ distribution} \\
 & $\log_{10}({\rm M}_{\star})$ & $z$ & $n$ & $U-B$ & $r_{e}$ & 
$\log_{10}({\rm M}_{\star})$ & $z$ & $n$ & $U-B$ & $r_{e}$
}
\startdata
\hline
$0.4 < z < 1.2$  & 
$-0.003$ & $-0.004$ & $0.047$ & 
$0.001$ & $0.543$ &  
\multirow{2}{*}{$-0.002$} & \multirow{2}{*}{$-0.003$} &
\multirow{2}{*}{$0.057$} & \multirow{2}{*}{$-0.001$} &
\multirow{2}{*}{$0.559$} \\
$10 < \log_{10}({\rm M}_{\star}) < 11$ & $\pm 0.030$ & $\pm
0.013$ &$\pm 0.248$
&$\pm 0.011$ &  $ \pm 0.220$ &   &  &  &  &  \\
\hline
$0.7 < z < 1.2$  & 
$-0.022$ & $-0.004$ & $0.023$ & 
$-0.008$ & $0.688$ &  
\multirow{2}{*}{$-0.005$} & \multirow{2}{*}{$-0.003$} &
\multirow{2}{*}{$0.075$} & \multirow{2}{*}{$-0.010$} &
\multirow{2}{*}{$0.632$} \\
$10 < \log_{10}({\rm M}_{\star}) < 11$ & $\pm 0.026$ & $\pm
0.008$ &$\pm 0.272$
&$\pm 0.012$ &  $ \pm 0.251$ &   &  &  &  &  \\
\hline
$0.4 < z < 1.05$  & 
$0.010$ & $-0.005$ & $0.085$ & 
$-0.005$ & $0.393$ &  
\multirow{2}{*}{$-0.003$} & \multirow{2}{*}{$-0.002$} &
\multirow{2}{*}{$0.078$} & \multirow{2}{*}{$0.001$} &
\multirow{2}{*}{$0.449$} \\
$10 < \log_{10}({\rm M}_{\star}) < 11$ & $\pm 0.024$ & $\pm
0.011$ &$\pm 0.226$
&$\pm 0.011$ &  $ \pm 0.209$ &   &  &  &  &  \\
\hline
$0.4 < z < 1.2$  & 
$-0.008$ & $0.001$ & $0.150$ & 
$0.010$ & $0.619$ &  
\multirow{2}{*}{$-0.003$} & \multirow{2}{*}{$-0.002$} &
\multirow{2}{*}{$0.085$} & \multirow{2}{*}{$0.012$} &
\multirow{2}{*}{$0.625$} \\
$10.5 < \log_{10}({\rm M}_{\star}) < 11$ & $\pm 0.017$ & $\pm
0.015$ &$\pm 0.285$
&$\pm 0.012$ &  $ \pm 0.274$ &   &  &  &  &  \\
\hline
$0.7 < z < 1.2$ & &  & &  & &  & &  & & \\ 
$10 < \log_{10}({\rm M}_{\star}) < 11$ & $-0.002$ & $-0.002$ & $0.145$
& $-0.010$ & $0.802$ &   
\multirow{3}{*}{$-0.002$} & \multirow{3}{*}{$-0.003$} &
\multirow{3}{*}{$0.090$} & \multirow{3}{*}{$-0.006$} &
\multirow{3}{*}{$0.826$} \\
$n > 2.5$ & $\pm 0.030$ & $\pm
0.009$ & $\pm 0.277$ & $\pm 0.013$ &  $\pm 0.239$ &   &  &  &  &  \\
\hline
$0.7 < z < 1.2$ & &  & &  & &  & &  & & \\ 
$10 < \log_{10}({\rm M}_{\star}) < 11$ & $-0.003$ & $-0.005$ & $-0.128$
& $-0.006$ & $0.909$ &   
\multirow{3}{*}{$-0.001$} & \multirow{3}{*}{$-0.006$} &
\multirow{3}{*}{$0.175$} & \multirow{3}{*}{$0.000$} &
\multirow{3}{*}{$0.981$} \\
$n > 2.5$,$b/a > 0.4$ & $\pm 0.030$ & $\pm
0.016$ & $\pm 0.301$ & $\pm 0.013$ &  $\pm 0.273$ &   &  &  &  &  \\
\hline
$0.7 < z < 1.2$ & &  & &  & &  & &  & & \\ 
$10 < \log_{10}({\rm M}_{\star}) < 11$ & $0.000$ & $0.001$ & $0.083$
& $-0.001$ & $0.426$ &   
\multirow{3}{*}{$0.000$} & \multirow{3}{*}{$0.000$} &
\multirow{3}{*}{$0.050$} & \multirow{3}{*}{$0.000$} &
\multirow{3}{*}{$0.471$} \\
$\mu < 23.5$ & $\pm 0.024$ & $\pm
0.011$ & $\pm 0.214$ & $\pm 0.010$ &  $\pm 0.224$ &   &  &  &  &  \\
\hline
\enddata
\tablecomments{For the same galaxy samples listed in Table
  \ref{res_tab1}, we list the mean and median of the distribution of
  the differences in the Hodges-Lehmann estimator of the mean stellar
  mass, redshift, S\'{e}rsic index, color, and size for the
  high-density samples relative to that for each of the low-density
  samples (see Figure \ref{fig_hlmeans}). Note that stellar masses and
  sizes are in units of $h^{-2}\ {\rm M}_{\sun}$ and $h^{-1}$ physical
  kpc, respectively. Finally, note that the samples listed here are
  matched in stellar mass, redshift, and S\'{e}rsic index, by
  construction.}
\end{deluxetable*}

\section{Discussion}
\label{sec_disc}

Previous efforts to study the environmental dependence of the
size-stellar mass relation at $z < 2$ within the early-type galaxy
population have been relatively few in number and have often found no
significant trends with local galaxy density. Comparing the
morphologies of massive galaxies in the low-redshift ($z=0.165$) Abell
901/902 supercluster to those of comparable field samples selected
from the Space Telescope A901/2 Galaxy Evolution Survey
\citep[STAGES,][]{gray09}, \citet{maltby10} detect no significant
relationship between galaxy structure and environment within the
early-type population. Focusing on galaxy groups identified in the
Sloan Digital Sky Survey \citep[SDSS,][]{york00}, a less extreme
subdivision of the environment distribution, \citet{guo09} also find
no significant evidence for a correlation between local environment
and the size or S\'{e}rsic index within the local early-type
population \citep[see also][]{weinmann09, nair10}. In contrast,
previous studies of brightest cluster galaxies (BCGs) in the local
Universe find that BCGs tend to be larger than early-types of
comparable stellar mass in less-dense environs \citep[e.g.,][see also
\citealt{bk06}]{bernardi07, vdl07, desroches07, liu08}. However, this
environmental dependence apparent in the BCG population is likely the
result of cluster-specific mechanisms that drive the formation of this
rare subset of the massive early-type galaxy population and may not be
indicative of the massive early-type population as a whole. We note
that our DEEP2/DEEP3 sample includes very few (if any) systems that
will evolve into BCGs at $z \sim 0$.

Beyond the local Universe, \citet{valentinuzzi10} find no significant
variation in the size-stellar mass relation for massive early-types in
clusters at $z \sim 0.7$ relative to that for comparable systems in
the field, using data drawn from the ESO Distant Clusters Survey
\citep[EDisCS,][]{white05}. At yet higher redshift, \citet{rettura10}
compare the sizes of massive ellipticals in an X-ray-luminous cluster
at $z = 1.237$ to those of correspondingly-massive systems in the
field, finding no significant variation in size with
environment. Utilizing the same data set, however, \citet{cimatti08}
propose (though without quantifying) a possible correlation between
size and environment similar in nature to that found within our
DEEP2/DEEP3 sample (that is, higher-density regions favoring
less-compact galaxies). Nevertheless, using largely the same galaxy
samples as these two previous studies, recent work from
\citet{raichoor11} argues for the opposite trend such that galaxies in
high-density regions are smaller than their field counterparts. The
significance of the measured correlation between size and environment,
however, is dramatically overstated by \citet{raichoor11}, with their
results actually consistent with no environment
dependence.\footnote{The average sizes for cluster, group, and field
  early-types as given in Table 1 of \citet{raichoor11} are all
  consistent at the $1\sigma$ level.} The lack of a significant
environment dependence reported in these previous studies is likely
due to [\emph{i}] the smaller sample sizes employed (the
\citet{rettura10} and \citet{cimatti08} analyses included a total of
$45$ ellipticals across all environments), [\emph{ii}] the use of
less-precise environment measures (e.g., relying on photometric
redshifts such that ``field'' or low-density samples can be strongly
contaminated by group members), and/or [\emph{iii}] differences in the
redshift range probed.

Our results, which show a significant size-density relation at fixed
stellar mass within the massive, red galaxy population, suggest that
the structural evolution of massive early-type systems occurs
preferentially in overdense environments (i.e., groups, given the lack
of massive clusters in our sample, \citealt{gerke05, gerke12}). This
environmental dependence is in general agreement with a model of
galaxy formation in which minor, ``dry'' mergers are a critical
mechanism in driving structural evolution at $z < 2$. Moreover, our
DEEP2/DEEP3 results also suggest that early-types in higher-density
regions evolved structurally prior to their counterparts in
low-density regions, growing from highly-compact systems at $z \sim
2$--$3$ to more extended systems at $z \sim 0$. The earlier onset of
this evolution in overdense environments is in agreement with studies
of stellar populations locally \citep[e.g.,][]{cooper10a} as well as
studies of the color-density relation at intermediate redshift
\citep[e.g.,][]{cooper06, cooper07, gerke07}, which find that galaxies
in high-density environments have typically ceased their star
formation earlier than those in less-dense environs. Studies of the
Fundamental Plane \citep[FP,][]{dd87, dressler87} support this picture
of accelerated evolution in high-density environments, with several
analyses finding that galaxies in high-density regions tend to reach
the FP more quickly than those in low-density regions
\citep{vandokkum01, gebhardt03, treu05, moran05}.

A correlation between size and environment at fixed stellar mass
within the massive early-type population is arguably in conflict with
scenarios in which the observed size evolution at intermediate
redshift is driven by quasar feedback, as quasars at $z \sim 1$ are
generally not found to reside in overdense environments, especially in
relation to the early-type galaxy population. Using cross-correlation
techniques and measurements of local environment analogous to those
presented herein, \citet{coil07} find that quasars at $z \sim 1$
cluster like blue galaxies, such that they favor regions of average
galaxy density ($\log_{10}(1+\delta_3) \sim 0$). While these results
suggest that quasars are unlikely to be responsible for the larger
sizes of early-types in high-density regions, it should be noted that
the clustering measurements of \citet{coil07} are roughly
$1$--$2\sigma$ lower than that of similar studies. For example,
\citet[][see also \citealt{porciani04,grazian04,myers06}]{croom05}
find that quasars at $0.3 < z < 2.2$ have a bias $\gtrsim \! 2\sigma$
higher than that found by \citet{coil07}, while \citet{serber06} find
an excess of $\sim \! L^{*}$ galaxies on $25$ kpc to $1$ Mpc ($h=0.7$)
projected comoving scales around quasars at $z < 0.4$ \citep[see
also][]{hennawi06, myers08}. For comparison, the median
third-nearest-neighbor distance for our sample of early-type galaxies
(over all environments) is $\sim \! 0.6$ $h^{-1}$ comoving Mpc in
projection.  In addition, \citet{croom05} find that the average dark
matter halo mass for quasars at intermediate redshift is roughly
consistent with (within a factor of a few of) the minimum halos mass
inferred for groups at $z \sim 1$ \citep[][see also
\citealt{hopkins07}]{coil06}.  Altogether, clustering and environment
studies do not support (though also do not clearly exclude) quasars as
a viable mechanism for size evolution at $z < 2$; regardless,
questions still remain as to how quasar activity would be fueled in a
massive early-type system at $z > 1$, as the standard scenario
involving the major merger of two gas-rich systems
\citep[e.g.,][]{springel05, dimatteo05, hopkins06} fails to accurately
describe massive early-type systems at $z \gtrsim 1$, which are
relatively gas-poor and have stellar populations with
luminosity-weighted ages of $> \!  1$--$2$ Gyr
\citep[e.g.,][]{daddi05, longhetti05, treu05b, schiavon06, combes07}.

In contrast to quasars, Seyfert galaxies at $z \sim 1$ typically
reside in higher-density environs, similar to those of galaxies on the
red sequence \citep[][but see also
\citealt{silverman09}]{georgakakis07, georgakakis08, coil09,
  bradshaw11, digby11}. In addition, systems exhibiting line ratios
consistent with Low Ionization Nuclear Emission-line Regions
\citep[LINERS,][]{heckman80} at $z \sim 1$, which tend to reside on
the red sequence, are likely to inhabit slightly more overdense
regions even relative to galaxies of like color and luminosity
\citep[i.e., stellar mass,][]{yan06, yan11, montero-dorta09, juneau11}
--- though, recent work suggests that LINERs may not be the product of
AGN activity \citep{yan11b}. While lower-luminosity AGN are found in
high-density regions at $z \sim 1$, consistent with being the driving
mechanism behind the observed size evolution of early-type galaxies,
the lack of variation in the outflow velocities of winds observed in
AGN hosts versus star-forming galaxies suggests that low-luminosity
AGN may not play a dominant role in galactic feedback \citep[][but see
also \citealt{hainline11}]{rupke05, weiner09, rubin10, rubin11,
  coil11}. Furthermore, the mass loss needed to produce a factor of
$\gtrsim \! 2$ increase in size is of order $30\%$--$50\%$
\citep{zhao02, hopkins10b}, beyond the expected impact of feedback
from lower-luminosity AGN or evolved stars \citep[e.g.,][]{damjanov09}.

\section{Summary}
\label{sec_summary}

Using data from the DEEP2 and DEEP3 Galaxy Redshift Surveys, we have
completed a detailed study of the relationship between galaxy
structure and environment on the massive ($10 < \log_{10}({\rm
  M}_{\star}/h^{-2}\ {\rm M}_{\sun}) < 11$) end of the red sequence at
intermediate redshift. Our principal result is that at fixed stellar
mass, redshift, S\'{e}rsic index, and rest-frame color we find a
significant relationship between galaxy size and local galaxy density
at $z \sim 0.75$, such that early-type galaxies in high-density
regions are more extended than their counterparts in low-density
environs. This result is robust to variations in the sample selection
procedure, including selection limits based on axis ratio, surface
brightness, and S\'{e}rsic index. The observed correlation between
size and environment is consistent with a scenario in which minor, dry
mergers play a critical role in the structural evolution of massive,
early-type galaxies at $z < 2$ and in which the evolution of massive
ellipticals is accelerated in high-density regions. Future work, for
example from the CANDELS {\it HST}/WFC3-IR imaging program, will soon
enable complementary analyses at yet higher redshift and in more
extreme environments such as massive clusters
\citep[e.g.,][]{papovich11}.

\vspace*{0.25in} 

\acknowledgments MCC acknowledges support for this work provided by
NASA through Hubble Fellowship grant \#HF-51269.01-A, awarded by the
Space Telescope Science Institute, which is operated by the
Association of Universities for Research in Astronomy, Inc., for NASA,
under contract NAS 5-26555. This work was also supported in part by
NSF grants AST-0507428, AST-0507483, AST-0071048, AST-0071198,
AST-0808133, and AST-0806732 as well as {\it Hubble Space Telescope}
Archival grant HST-AR-10947.01 and NASA grant
HST-G0-10134.13-A. Additional support was provided by NASA through the
Spitzer Space Telescope Fellowship Program. MCC acknowledges support
from the Southern California Center for Galaxy Evolution, a
multi-campus research program funded by the University of California
Office of Research. MCC thanks Mike Boylan-Kolchin for
helpful discussions in preparing this manuscript and also thanks Greg
Wirth and the entire Keck Observatory staff for their help in the
acquisition of the DEEP2 and DEEP3 Keck/DEIMOS data. Finally, MCC
thanks the anonymous referee for their insightful comments and
suggestions that improved this work.

We also wish to recognize and acknowledge the highly significant
cultural role and reverence that the summit of Mauna Kea has always
had within the indigenous Hawaiian community. It is a privilege to be
given the opportunity to conduct observations from this mountain.

{\it Facilities:} \facility{Keck:II (DEIMOS)}, \facility{HST (ACS)}


\begin{thebibliography}{144}
\expandafter\ifx\csname natexlab\endcsname\relax\def\natexlab#1{#1}\fi

\bibitem[{{Baldry} {et~al.}(2006){Baldry}, {Balogh}, {Bower}, {Glazebrook},
  {Nichol}, {Bamford}, \& {Budavari}}]{baldry06}
{Baldry}, I.~K., {Balogh}, M.~L., {Bower}, R.~G., {Glazebrook}, K., {Nichol},
  R.~C., {Bamford}, S.~P., \& {Budavari}, T. 2006, \mnras, 373, 469

\bibitem[{{Balogh} {et~al.}(2001){Balogh}, {Christlein}, {Zabludoff}, \&
  {Zaritsky}}]{balogh01}
{Balogh}, M.~L., {Christlein}, D., {Zabludoff}, A.~I., \& {Zaritsky}, D. 2001,
  \apj, 557, 117

\bibitem[{{Bell} {et~al.}(2003){Bell}, {McIntosh}, {Katz}, \&
  {Weinberg}}]{bell03}
{Bell}, E.~F., {McIntosh}, D.~H., {Katz}, N., \& {Weinberg}, M.~D. 2003, \apjs,
  149, 289

\bibitem[{{Bernardi} {et~al.}(2007){Bernardi}, {Hyde}, {Sheth}, {Miller}, \&
  {Nichol}}]{bernardi07}
{Bernardi}, M., {Hyde}, J.~B., {Sheth}, R.~K., {Miller}, C.~J., \& {Nichol},
  R.~C. 2007, \aj, 133, 1741

\bibitem[{{Bezanson} {et~al.}(2009){Bezanson}, {van Dokkum}, {Tal},
  {Marchesini}, {Kriek}, {Franx}, \& {Coppi}}]{bezanson09}
{Bezanson}, R., {van Dokkum}, P.~G., {Tal}, T., {Marchesini}, D., {Kriek}, M.,
  {Franx}, M., \& {Coppi}, P. 2009, \apj, 697, 1290

\bibitem[{{Blanton} {et~al.}(2005){Blanton}, {Eisenstein}, {Hogg}, {Schlegel},
  \& {Brinkmann}}]{blanton05}
{Blanton}, M.~R., {Eisenstein}, D., {Hogg}, D.~W., {Schlegel}, D.~J., \&
  {Brinkmann}, J. 2005, \apj, 629, 143

\bibitem[{{Bolzonella} {et~al.}(2010)}]{bolzonella10}
{Bolzonella}, M. {et~al.} 2010, \aap, 524, A76+

\bibitem[{{Bournaud} {et~al.}(2007){Bournaud}, {Jog}, \& {Combes}}]{bournaud07}
{Bournaud}, F., {Jog}, C.~J., \& {Combes}, F. 2007, \aap, 476, 1179

\bibitem[{{Boylan-Kolchin} \& {Ma}(2007)}]{bk07}
{Boylan-Kolchin}, M. \& {Ma}, C.-P. 2007, \mnras, 374, 1227

\bibitem[{{Boylan-Kolchin} {et~al.}(2006){Boylan-Kolchin}, {Ma}, \&
  {Quataert}}]{bk06}
{Boylan-Kolchin}, M., {Ma}, C.-P., \& {Quataert}, E. 2006, \mnras, 369, 1081

\bibitem[{{Bradshaw} {et~al.}(2011)}]{bradshaw11}
{Bradshaw}, E.~J. {et~al.} 2011, \mnras, 415, 2626

\bibitem[{{Bundy} {et~al.}(2005){Bundy}, {Ellis}, \& {Conselice}}]{bundy05}
{Bundy}, K., {Ellis}, R.~S., \& {Conselice}, C.~J. 2005, \apj, 625, 621

\bibitem[{{Bundy} {et~al.}(2006)}]{bundy06}
{Bundy}, K. {et~al.} 2006, \apj, 651, 120

\bibitem[{{Bundy} {et~al.}(2010)}]{bundy10}
---. 2010, \apj, 719, 1969

\bibitem[{{Capak} {et~al.}(2007){Capak}, {Abraham}, {Ellis}, {Mobasher},
  {Scoville}, {Sheth}, \& {Koekemoer}}]{capak07}
{Capak}, P., {Abraham}, R.~G., {Ellis}, R.~S., {Mobasher}, B., {Scoville}, N.,
  {Sheth}, K., \& {Koekemoer}, A. 2007, \apjs, 172, 284

\bibitem[{{Cavaliere} {et~al.}(1992){Cavaliere}, {Colafrancesco}, \&
  {Menci}}]{cavaliere92}
{Cavaliere}, A., {Colafrancesco}, S., \& {Menci}, N. 1992, \apj, 392, 41

\bibitem[{{Cenarro} \& {Trujillo}(2009)}]{cenarro09}
{Cenarro}, A.~J. \& {Trujillo}, I. 2009, \apjl, 696, L43

\bibitem[{{Cheng} {et~al.}(2011){Cheng}, {Faber}, {Simard}, {Graves}, {Lopez},
  {Yan}, \& {Cooper}}]{cheng11}
{Cheng}, J.~Y., {Faber}, S.~M., {Simard}, L., {Graves}, G.~J., {Lopez}, E.~D.,
  {Yan}, R., \& {Cooper}, M.~C. 2011, \mnras, 412, 727

\bibitem[{{Cimatti} {et~al.}(2008)}]{cimatti08}
{Cimatti}, A. {et~al.} 2008, \aap, 482, 21

\bibitem[{{Coil} {et~al.}(2007){Coil}, {Hennawi}, {Newman}, {Cooper}, \&
  {Davis}}]{coil07}
{Coil}, A.~L., {Hennawi}, J.~F., {Newman}, J.~A., {Cooper}, M.~C., \& {Davis},
  M. 2007, \apj, 654, 115

\bibitem[{{Coil} {et~al.}(2004){Coil}, {Newman}, {Kaiser}, {Davis}, {Ma},
  {Kocevski}, \& {Koo}}]{coil04b}
{Coil}, A.~L., {Newman}, J.~A., {Kaiser}, N., {Davis}, M., {Ma}, C.-P.,
  {Kocevski}, D.~D., \& {Koo}, D.~C. 2004, \apj, 617, 765

\bibitem[{{Coil} {et~al.}(2011){Coil}, {Weiner}, {Holz}, {Cooper}, {Yan}, \&
  {Aird}}]{coil11}
{Coil}, A.~L., {Weiner}, B.~J., {Holz}, D.~E., {Cooper}, M.~C., {Yan}, R., \&
  {Aird}, J. 2011, ArXiv e-prints

\bibitem[{{Coil} {et~al.}(2006)}]{coil06}
{Coil}, A.~L. {et~al.} 2006, \apj, 638, 668

\bibitem[{{Coil} {et~al.}(2008)}]{coil08}
---. 2008, \apj, 672, 153

\bibitem[{{Coil} {et~al.}(2009)}]{coil09}
---. 2009, \apj, 701, 1484

\bibitem[{{Combes} {et~al.}(2007){Combes}, {Young}, \& {Bureau}}]{combes07}
{Combes}, F., {Young}, L.~M., \& {Bureau}, M. 2007, \mnras, 377, 1795

\bibitem[{{Conselice} {et~al.}(2008){Conselice}, {Bundy}, {U}, {Eisenhardt},
  {Lotz}, \& {Newman}}]{conselice08}
{Conselice}, C.~J., {Bundy}, K., {U}, V., {Eisenhardt}, P., {Lotz}, J., \&
  {Newman}, J. 2008, \mnras, 383, 1366

\bibitem[{{Cooper} {et~al.}(2010{\natexlab{a}}){Cooper}, {Gallazzi}, {Newman},
  \& {Yan}}]{cooper10a}
{Cooper}, M.~C., {Gallazzi}, A., {Newman}, J.~A., \& {Yan}, R.
  2010{\natexlab{a}}, \mnras, 402, 1942

\bibitem[{{Cooper} {et~al.}(2005){Cooper}, {Newman}, {Madgwick}, {Gerke},
  {Yan}, \& {Davis}}]{cooper05}
{Cooper}, M.~C., {Newman}, J.~A., {Madgwick}, D.~S., {Gerke}, B.~F., {Yan}, R.,
  \& {Davis}, M. 2005, \apj, 634, 833

\bibitem[{{Cooper} {et~al.}(2006)}]{cooper06}
{Cooper}, M.~C. {et~al.} 2006, \mnras, 370, 198

\bibitem[{{Cooper} {et~al.}(2007)}]{cooper07}
---. 2007, \mnras, 376, 1445

\bibitem[{{Cooper} {et~al.}(2008)}]{cooper08}
---. 2008, \mnras, 383, 1058

\bibitem[{{Cooper} {et~al.}(2010{\natexlab{b}})}]{cooper10b}
---. 2010{\natexlab{b}}, \mnras, 409, 337

\bibitem[{{Cooper} {et~al.}(2011)}]{cooper11}
---. 2011, \apjs, 193, 14

\bibitem[{{Croom} {et~al.}(2005)}]{croom05}
{Croom}, S.~M. {et~al.} 2005, \mnras, 356, 415

\bibitem[{{Croton} {et~al.}(2005)}]{croton05}
{Croton}, D.~J. {et~al.} 2005, \mnras, 356, 1155

\bibitem[{{Daddi} {et~al.}(2005)}]{daddi05}
{Daddi}, E. {et~al.} 2005, \apj, 626, 680

\bibitem[{{Damjanov} {et~al.}(2009)}]{damjanov09}
{Damjanov}, I. {et~al.} 2009, \apj, 695, 101

\bibitem[{{Damjanov} {et~al.}(2011)}]{damjanov11}
---. 2011, \apjl, 739, L44+

\bibitem[{{Darg} {et~al.}(2010)}]{darg10}
{Darg}, D.~W. {et~al.} 2010, \mnras, 401, 1552

\bibitem[{{Davis} {et~al.}(2003)}]{davis03}
{Davis}, M. {et~al.} 2003, in Society of Photo-Optical Instrumentation
  Engineers (SPIE) Conference Series, Vol. 4834, Society of Photo-Optical
  Instrumentation Engineers (SPIE) Conference Series, ed. {P.~Guhathakurta},
  161--172

\bibitem[{{Davis} {et~al.}(2007)}]{davis07}
{Davis}, M. {et~al.} 2007, \apjl, 660, L1

\bibitem[{{Desroches} {et~al.}(2007){Desroches}, {Quataert}, {Ma}, \&
  {West}}]{desroches07}
{Desroches}, L.-B., {Quataert}, E., {Ma}, C.-P., \& {West}, A.~A. 2007, \mnras,
  377, 402

\bibitem[{{Di Matteo} {et~al.}(2005){Di Matteo}, {Springel}, \&
  {Hernquist}}]{dimatteo05}
{Di Matteo}, T., {Springel}, V., \& {Hernquist}, L. 2005, \nat, 433, 604

\bibitem[{{Digby-North} {et~al.}(2011)}]{digby11}
{Digby-North}, J.~A. {et~al.} 2011, submitted

\bibitem[{{Djorgovski} \& {Davis}(1987)}]{dd87}
{Djorgovski}, S. \& {Davis}, M. 1987, \apj, 313, 59

\bibitem[{{Dressler} {et~al.}(1987){Dressler}, {Lynden-Bell}, {Burstein},
  {Davies}, {Faber}, {Terlevich}, \& {Wegner}}]{dressler87}
{Dressler}, A., {Lynden-Bell}, D., {Burstein}, D., {Davies}, R.~L., {Faber},
  S.~M., {Terlevich}, R., \& {Wegner}, G. 1987, \apj, 313, 42

\bibitem[{{Efron}(1981)}]{efron81}
{Efron}, B. 1981, Biometrika, 68, 589

\bibitem[{{Elbaz} {et~al.}(2007)}]{elbaz07}
{Elbaz}, D. {et~al.} 2007, \aap, 468, 33

\bibitem[{{Fakhouri} \& {Ma}(2009)}]{fakhouri09}
{Fakhouri}, O. \& {Ma}, C.-P. 2009, \mnras, 394, 1825

\bibitem[{{Fan} {et~al.}(2010){Fan}, {Lapi}, {Bressan}, {Bernardi}, {De Zotti},
  \& {Danese}}]{fan10}
{Fan}, L., {Lapi}, A., {Bressan}, A., {Bernardi}, M., {De Zotti}, G., \&
  {Danese}, L. 2010, \apj, 718, 1460

\bibitem[{{Fan} {et~al.}(2008){Fan}, {Lapi}, {De Zotti}, \& {Danese}}]{fan08}
{Fan}, L., {Lapi}, A., {De Zotti}, G., \& {Danese}, L. 2008, \apjl, 689, L101

\bibitem[{{Gebhardt} {et~al.}(2003){Gebhardt}, {Faber}, {Koo}, {Im}, {Simard},
  {Illingworth}, {Phillips}, {Sarajedini}, {Vogt}, {Weiner}, \&
  {Willmer}}]{gebhardt03}
{Gebhardt}, K., {Faber}, S.~M., {Koo}, D.~C., {Im}, M., {Simard}, L.,
  {Illingworth}, G.~D., {Phillips}, A.~C., {Sarajedini}, V.~L., {Vogt}, N.~P.,
  {Weiner}, B., \& {Willmer}, C.~N.~A. 2003, \apj, 597, 239

\bibitem[{{Georgakakis} {et~al.}(2008){Georgakakis}, {Gerke}, {Nandra},
  {Laird}, {Coil}, {Cooper}, \& {Newman}}]{georgakakis08}
{Georgakakis}, A., {Gerke}, B.~F., {Nandra}, K., {Laird}, E.~S., {Coil}, A.~L.,
  {Cooper}, M.~C., \& {Newman}, J.~A. 2008, \mnras, 391, 183

\bibitem[{{Georgakakis} {et~al.}(2007)}]{georgakakis07}
{Georgakakis}, A. {et~al.} 2007, \apjl, 660, L15

\bibitem[{{Gerke} {et~al.}(2005)}]{gerke05}
{Gerke}, B.~F. {et~al.} 2005, \apj, 625, 6

\bibitem[{{Gerke} {et~al.}(2007)}]{gerke07}
---. 2007, \mnras, 376, 1425

\bibitem[{{Gerke} {et~al.}(2012)}]{gerke12}
---. 2012, submitted

\bibitem[{{Giavalisco} {et~al.}(2004)}]{giavalisco04}
{Giavalisco}, M. {et~al.} 2004, \apjl, 600, L93

\bibitem[{{Gray} {et~al.}(2009)}]{gray09}
{Gray}, M.~E. {et~al.} 2009, \mnras, 393, 1275

\bibitem[{{Grazian} {et~al.}(2004){Grazian}, {Negrello}, {Moscardini},
  {Cristiani}, {Haehnelt}, {Matarrese}, {Omizzolo}, \& {Vanzella}}]{grazian04}
{Grazian}, A., {Negrello}, M., {Moscardini}, L., {Cristiani}, S., {Haehnelt},
  M.~G., {Matarrese}, S., {Omizzolo}, A., \& {Vanzella}, E. 2004, \aj, 127, 592

\bibitem[{{Grogin} {et~al.}(2011)}]{grogin11}
{Grogin}, N.~A. {et~al.} 2011, ArXiv e-prints

\bibitem[{{Guo} {et~al.}(2009)}]{guo09}
{Guo}, Y. {et~al.} 2009, \mnras, 398, 1129

\bibitem[{{Hainline} {et~al.}(2011){Hainline}, {Shapley}, {Greene}, \&
  {Steidel}}]{hainline11}
{Hainline}, K.~N., {Shapley}, A.~E., {Greene}, J.~E., \& {Steidel}, C.~C. 2011,
  \apj, 733, 31

\bibitem[{{H{\"a}u{\ss}ler} {et~al.}(2011){H{\"a}u{\ss}ler}, {Barden},
  {Bamford}, \& {Rojas}}]{haussler11}
{H{\"a}u{\ss}ler}, B., {Barden}, M., {Bamford}, S.~P., \& {Rojas}, A. 2011, in
  Astronomical Society of the Pacific Conference Series, Vol. 442, Astronomical
  Society of the Pacific Conference Series, ed. {I.~N.~Evans, A.~Accomazzi,
  D.~J.~Mink, \& A.~H.~Rots}, 155--+

\bibitem[{{H{\"a}u{\ss}ler} {et~al.}(2007)}]{haussler07}
{H{\"a}u{\ss}ler}, B. {et~al.} 2007, \apjs, 172, 615

\bibitem[{{Heckman}(1980)}]{heckman80}
{Heckman}, T.~M. 1980, \aap, 87, 152

\bibitem[{{Hennawi} {et~al.}(2006)}]{hennawi06}
{Hennawi}, J.~F. {et~al.} 2006, \aj, 131, 1

\bibitem[{{Hodges} \& {Lehmann}(1963)}]{hodges63}
{Hodges}, J.~R. \& {Lehmann}, E.~L. 1963, The Annals of Mathematical
  Statistics, 34, 598

\bibitem[{{Hopkins} {et~al.}(2010{\natexlab{a}}){Hopkins}, {Bundy},
  {Hernquist}, {Wuyts}, \& {Cox}}]{hopkins10b}
{Hopkins}, P.~F., {Bundy}, K., {Hernquist}, L., {Wuyts}, S., \& {Cox}, T.~J.
  2010{\natexlab{a}}, \mnras, 401, 1099

\bibitem[{{Hopkins} {et~al.}(2009{\natexlab{a}}){Hopkins}, {Bundy}, {Murray},
  {Quataert}, {Lauer}, \& {Ma}}]{hopkins09b}
{Hopkins}, P.~F., {Bundy}, K., {Murray}, N., {Quataert}, E., {Lauer}, T.~R., \&
  {Ma}, C.-P. 2009{\natexlab{a}}, \mnras, 398, 898

\bibitem[{{Hopkins} {et~al.}(2006){Hopkins}, {Hernquist}, {Cox}, {Di Matteo},
  {Robertson}, \& {Springel}}]{hopkins06}
{Hopkins}, P.~F., {Hernquist}, L., {Cox}, T.~J., {Di Matteo}, T., {Robertson},
  B., \& {Springel}, V. 2006, \apjs, 163, 1

\bibitem[{{Hopkins} {et~al.}(2009{\natexlab{b}}){Hopkins}, {Hernquist}, {Cox},
  {Keres}, \& {Wuyts}}]{hopkins09a}
{Hopkins}, P.~F., {Hernquist}, L., {Cox}, T.~J., {Keres}, D., \& {Wuyts}, S.
  2009{\natexlab{b}}, \apj, 691, 1424

\bibitem[{{Hopkins} {et~al.}(2007){Hopkins}, {Lidz}, {Hernquist}, {Coil},
  {Myers}, {Cox}, \& {Spergel}}]{hopkins07}
{Hopkins}, P.~F., {Lidz}, A., {Hernquist}, L., {Coil}, A.~L., {Myers}, A.~D.,
  {Cox}, T.~J., \& {Spergel}, D.~N. 2007, \apj, 662, 110

\bibitem[{{Hopkins} {et~al.}(2010{\natexlab{b}}){Hopkins}, {Murray},
  {Quataert}, \& {Thompson}}]{hopkins10a}
{Hopkins}, P.~F., {Murray}, N., {Quataert}, E., \& {Thompson}, T.~A.
  2010{\natexlab{b}}, \mnras, 401, L19

\bibitem[{{Juneau} {et~al.}(2011){Juneau}, {Dickinson}, {Alexander}, \&
  {Salim}}]{juneau11}
{Juneau}, S., {Dickinson}, M., {Alexander}, D.~M., \& {Salim}, S. 2011, \apj,
  736, 104

\bibitem[{{Kauffmann} {et~al.}(2004){Kauffmann}, {White}, {Heckman},
  {M{\'e}nard}, {Brinchmann}, {Charlot}, {Tremonti}, \&
  {Brinkmann}}]{kauffmann04}
{Kauffmann}, G., {White}, S.~D.~M., {Heckman}, T.~M., {M{\'e}nard}, B.,
  {Brinchmann}, J., {Charlot}, S., {Tremonti}, C., \& {Brinkmann}, J. 2004,
  \mnras, 353, 713

\bibitem[{{Khochfar} \& {Silk}(2006)}]{khochfar06}
{Khochfar}, S. \& {Silk}, J. 2006, \apjl, 648, L21

\bibitem[{{Kova{\v c}} {et~al.}(2010)}]{kovac10}
{Kova{\v c}}, K. {et~al.} 2010, \apj, 718, 86

\bibitem[{{Kriek} {et~al.}(2006)}]{kriek06}
{Kriek}, M. {et~al.} 2006, \apjl, 649, L71

\bibitem[{{Labb{\'e}} {et~al.}(2005)}]{labbe05}
{Labb{\'e}}, I. {et~al.} 2005, \apjl, 624, L81

\bibitem[{{Lin} {et~al.}(2007)}]{lin07}
{Lin}, L. {et~al.} 2007, \apjl, 660, L51

\bibitem[{{Lin} {et~al.}(2010)}]{lin10}
---. 2010, \apj, 718, 1158

\bibitem[{{Liu} {et~al.}(2008){Liu}, {Xia}, {Mao}, {Wu}, \& {Deng}}]{liu08}
{Liu}, F.~S., {Xia}, X.~Y., {Mao}, S., {Wu}, H., \& {Deng}, Z.~G. 2008, \mnras,
  385, 23

\bibitem[{{Longhetti} {et~al.}(2005)}]{longhetti05}
{Longhetti}, M. {et~al.} 2005, \mnras, 361, 897

\bibitem[{{Lotz} {et~al.}(2008)}]{lotz08}
{Lotz}, J.~M. {et~al.} 2008, \apj, 672, 177

\bibitem[{{Maltby} {et~al.}(2010)}]{maltby10}
{Maltby}, D.~T. {et~al.} 2010, \mnras, 402, 282

\bibitem[{{Mann} \& {Whitney}(1947)}]{mann47}
{Mann}, H.~B. \& {Whitney}, D.~R. 1947, The Annals of Mathematical Statistics,
  18, 50

\bibitem[{{McIntosh} {et~al.}(2008){McIntosh}, {Guo}, {Hertzberg}, {Katz},
  {Mo}, {van den Bosch}, \& {Yang}}]{mcintosh08}
{McIntosh}, D.~H., {Guo}, Y., {Hertzberg}, J., {Katz}, N., {Mo}, H.~J., {van
  den Bosch}, F.~C., \& {Yang}, X. 2008, \mnras, 388, 1537

\bibitem[{{Montero-Dorta} {et~al.}(2009)}]{montero-dorta09}
{Montero-Dorta}, A.~D. {et~al.} 2009, \mnras, 392, 125

\bibitem[{{Moran} {et~al.}(2005){Moran}, {Ellis}, {Treu}, {Smail}, {Dressler},
  {Coil}, \& {Smith}}]{moran05}
{Moran}, S.~M., {Ellis}, R.~S., {Treu}, T., {Smail}, I., {Dressler}, A.,
  {Coil}, A.~L., \& {Smith}, G.~P. 2005, \apj, 634, 977

\bibitem[{{Myers} {et~al.}(2008){Myers}, {Richards}, {Brunner}, {Schneider},
  {Strand}, {Hall}, {Blomquist}, \& {York}}]{myers08}
{Myers}, A.~D., {Richards}, G.~T., {Brunner}, R.~J., {Schneider}, D.~P.,
  {Strand}, N.~E., {Hall}, P.~B., {Blomquist}, J.~A., \& {York}, D.~G. 2008,
  \apj, 678, 635

\bibitem[{{Myers} {et~al.}(2006)}]{myers06}
{Myers}, A.~D. {et~al.} 2006, \apj, 638, 622

\bibitem[{{Naab} {et~al.}(2009){Naab}, {Johansson}, \& {Ostriker}}]{naab09}
{Naab}, T., {Johansson}, P.~H., \& {Ostriker}, J.~P. 2009, \apjl, 699, L178

\bibitem[{{Naab} {et~al.}(2007){Naab}, {Johansson}, {Ostriker}, \&
  {Efstathiou}}]{naab07}
{Naab}, T., {Johansson}, P.~H., {Ostriker}, J.~P., \& {Efstathiou}, G. 2007,
  \apj, 658, 710

\bibitem[{{Naab} {et~al.}(2006){Naab}, {Khochfar}, \& {Burkert}}]{naab06}
{Naab}, T., {Khochfar}, S., \& {Burkert}, A. 2006, \apjl, 636, L81

\bibitem[{{Nair} {et~al.}(2010){Nair}, {van den Bergh}, \& {Abraham}}]{nair10}
{Nair}, P.~B., {van den Bergh}, S., \& {Abraham}, R.~G. 2010, \apj, 715, 606

\bibitem[{{Newman} {et~al.}(2012)}]{newman12}
{Newman}, J.~A. {et~al.} 2012, in prep

\bibitem[{{Nipoti} {et~al.}(2009){Nipoti}, {Treu}, {Auger}, \&
  {Bolton}}]{nipoti09}
{Nipoti}, C., {Treu}, T., {Auger}, M.~W., \& {Bolton}, A.~S. 2009, \apjl, 706,
  L86

\bibitem[{{Oke} \& {Gunn}(1983)}]{oke83}
{Oke}, J.~B. \& {Gunn}, J.~E. 1983, \apj, 266, 713

\bibitem[{{Papovich} {et~al.}(2006)}]{papovich06}
{Papovich}, C. {et~al.} 2006, \apj, 640, 92

\bibitem[{{Papovich} {et~al.}(2011)}]{papovich11}
---. 2011, in prep

\bibitem[{{Peng} {et~al.}(2002){Peng}, {Ho}, {Impey}, \& {Rix}}]{peng02}
{Peng}, C.~Y., {Ho}, L.~C., {Impey}, C.~D., \& {Rix}, H.-W. 2002, \aj, 124, 266

\bibitem[{{Peng} {et~al.}(2010){Peng}, {Ho}, {Impey}, \& {Rix}}]{peng10}
---. 2010, \aj, 139, 2097

\bibitem[{{Porciani} {et~al.}(2004){Porciani}, {Magliocchetti}, \&
  {Norberg}}]{porciani04}
{Porciani}, C., {Magliocchetti}, M., \& {Norberg}, P. 2004, \mnras, 355, 1010

\bibitem[{{Press} {et~al.}(1986){Press}, {Flannery}, \& {Teukolsky}}]{press86}
{Press}, W.~H., {Flannery}, B.~P., \& {Teukolsky}, S.~A. 1986, {Numerical
  recipes. The art of scientific computing}, ed. {Press, W.~H., Flannery,
  B.~P., \& Teukolsky, S.~A.}

\bibitem[{{Raichoor} {et~al.}(2011)}]{raichoor11}
{Raichoor}, A. {et~al.} 2011, ArXiv e-prints

\bibitem[{{Rettura} {et~al.}(2010)}]{rettura10}
{Rettura}, A. {et~al.} 2010, \apj, 709, 512

\bibitem[{{Rubin} {et~al.}(2011){Rubin}, {Prochaska}, {M{\'e}nard}, {Murray},
  {Kasen}, {Koo}, \& {Phillips}}]{rubin11}
{Rubin}, K.~H.~R., {Prochaska}, J.~X., {M{\'e}nard}, B., {Murray}, N., {Kasen},
  D., {Koo}, D.~C., \& {Phillips}, A.~C. 2011, \apj, 728, 55

\bibitem[{{Rubin} {et~al.}(2010){Rubin}, {Weiner}, {Koo}, {Martin},
  {Prochaska}, {Coil}, \& {Newman}}]{rubin10}
{Rubin}, K.~H.~R., {Weiner}, B.~J., {Koo}, D.~C., {Martin}, C.~L., {Prochaska},
  J.~X., {Coil}, A.~L., \& {Newman}, J.~A. 2010, \apj, 719, 1503

\bibitem[{{Rudnick} {et~al.}(2009)}]{rudnick09}
{Rudnick}, G. {et~al.} 2009, \apj, 700, 1559

\bibitem[{{Rupke} {et~al.}(2005){Rupke}, {Veilleux}, \& {Sanders}}]{rupke05}
{Rupke}, D.~S., {Veilleux}, S., \& {Sanders}, D.~B. 2005, \apj, 632, 751

\bibitem[{{Schiavon} {et~al.}(2006)}]{schiavon06}
{Schiavon}, R.~P. {et~al.} 2006, \apjl, 651, L93

\bibitem[{{Serber} {et~al.}(2006){Serber}, {Bahcall}, {M{\'e}nard}, \&
  {Richards}}]{serber06}
{Serber}, W., {Bahcall}, N., {M{\'e}nard}, B., \& {Richards}, G. 2006, \apj,
  643, 68

\bibitem[{{S\'{e}rsic}(1968)}]{sersic68}
{S\'{e}rsic}, J.~L. 1968, {Atlas de galaxias australes}, ed. {Sersic, J.~L.}

\bibitem[{{Silverman} {et~al.}(2009)}]{silverman09}
{Silverman}, J.~D. {et~al.} 2009, \apj, 695, 171

\bibitem[{{Springel} {et~al.}(2005){Springel}, {Di Matteo}, \&
  {Hernquist}}]{springel05}
{Springel}, V., {Di Matteo}, T., \& {Hernquist}, L. 2005, \apjl, 620, L79

\bibitem[{{Taylor} {et~al.}(2010){Taylor}, {Franx}, {Glazebrook}, {Brinchmann},
  {van der Wel}, \& {van Dokkum}}]{taylor10}
{Taylor}, E.~N., {Franx}, M., {Glazebrook}, K., {Brinchmann}, J., {van der
  Wel}, A., \& {van Dokkum}, P.~G. 2010, \apj, 720, 723

\bibitem[{{Toft} {et~al.}(2009){Toft}, {Franx}, {van Dokkum}, {F{\"o}rster
  Schreiber}, {Labbe}, {Wuyts}, \& {Marchesini}}]{toft09}
{Toft}, S., {Franx}, M., {van Dokkum}, P., {F{\"o}rster Schreiber}, N.~M.,
  {Labbe}, I., {Wuyts}, S., \& {Marchesini}, D. 2009, \apj, 705, 255

\bibitem[{{Treu} {et~al.}(2005{\natexlab{a}}){Treu}, {Ellis}, {Liao}, \& {van
  Dokkum}}]{treu05}
{Treu}, T., {Ellis}, R.~S., {Liao}, T.~X., \& {van Dokkum}, P.~G.
  2005{\natexlab{a}}, \apjl, 622, L5

\bibitem[{{Treu} {et~al.}(2005{\natexlab{b}}){Treu}, {Ellis}, {Liao}, {van
  Dokkum}, {Tozzi}, {Coil}, {Newman}, {Cooper}, \& {Davis}}]{treu05b}
{Treu}, T., {Ellis}, R.~S., {Liao}, T.~X., {van Dokkum}, P.~G., {Tozzi}, P.,
  {Coil}, A., {Newman}, J., {Cooper}, M.~C., \& {Davis}, M. 2005{\natexlab{b}},
  \apj, 633, 174

\bibitem[{{Trujillo} {et~al.}(2007){Trujillo}, {Conselice}, {Bundy}, {Cooper},
  {Eisenhardt}, \& {Ellis}}]{trujillo07}
{Trujillo}, I., {Conselice}, C.~J., {Bundy}, K., {Cooper}, M.~C., {Eisenhardt},
  P., \& {Ellis}, R.~S. 2007, \mnras, 382, 109

\bibitem[{{Trujillo} {et~al.}(2011){Trujillo}, {Ferreras}, \& {de La
  Rosa}}]{trujillo11}
{Trujillo}, I., {Ferreras}, I., \& {de La Rosa}, I.~G. 2011, \mnras, 415, 3903

\bibitem[{{Trujillo} {et~al.}(2006)}]{trujillo06}
{Trujillo}, I. {et~al.} 2006, \apj, 650, 18

\bibitem[{{Valentinuzzi} {et~al.}(2010){Valentinuzzi}, {Poggianti}, {Saglia},
  {Arag{\'o}n-Salamanca}, {Simard}, {S{\'a}nchez-Bl{\'a}zquez}, {D'onofrio},
  {Cava}, {Couch}, {Fritz}, {Moretti}, \& {Vulcani}}]{valentinuzzi10}
{Valentinuzzi}, T., {Poggianti}, B.~M., {Saglia}, R.~P.,
  {Arag{\'o}n-Salamanca}, A., {Simard}, L., {S{\'a}nchez-Bl{\'a}zquez}, P.,
  {D'onofrio}, M., {Cava}, A., {Couch}, W.~J., {Fritz}, J., {Moretti}, A., \&
  {Vulcani}, B. 2010, \apjl, 721, L19

\bibitem[{{van der Wel} {et~al.}(2009){van der Wel}, {Bell}, {van den Bosch},
  {Gallazzi}, \& {Rix}}]{vdw09}
{van der Wel}, A., {Bell}, E.~F., {van den Bosch}, F.~C., {Gallazzi}, A., \&
  {Rix}, H.-W. 2009, \apj, 698, 1232

\bibitem[{{van der Wel} {et~al.}(2007)}]{vdw07}
{van der Wel}, A. {et~al.} 2007, \apj, 670, 206

\bibitem[{{van Dokkum} {et~al.}(2001){van Dokkum}, {Franx}, {Kelson}, \&
  {Illingworth}}]{vandokkum01}
{van Dokkum}, P.~G., {Franx}, M., {Kelson}, D.~D., \& {Illingworth}, G.~D.
  2001, \apjl, 553, L39

\bibitem[{{van Dokkum} {et~al.}(2008)}]{vandokkum08}
{van Dokkum}, P.~G. {et~al.} 2008, \apjl, 677, L5

\bibitem[{{van Dokkum} {et~al.}(2010)}]{vandokkum10}
---. 2010, \apj, 709, 1018

\bibitem[{{von der Linden} {et~al.}(2007){von der Linden}, {Best}, {Kauffmann},
  \& {White}}]{vdl07}
{von der Linden}, A., {Best}, P.~N., {Kauffmann}, G., \& {White}, S.~D.~M.
  2007, \mnras, 379, 867

\bibitem[{{Wall} \& {Jenkins}(2003)}]{wall03}
{Wall}, J.~V. \& {Jenkins}, C.~R. 2003, {Practical Statistics for Astronomers}
  (Princeton Series in Astrophysics)

\bibitem[{{Weiner} {et~al.}(2009)}]{weiner09}
{Weiner}, B.~J. {et~al.} 2009, \apj, 692, 187

\bibitem[{{Weinmann} {et~al.}(2009){Weinmann}, {Kauffmann}, {van den Bosch},
  {Pasquali}, {McIntosh}, {Mo}, {Yang}, \& {Guo}}]{weinmann09}
{Weinmann}, S.~M., {Kauffmann}, G., {van den Bosch}, F.~C., {Pasquali}, A.,
  {McIntosh}, D.~H., {Mo}, H., {Yang}, X., \& {Guo}, Y. 2009, \mnras, 394, 1213

\bibitem[{{Wetzel} {et~al.}(2008){Wetzel}, {Schulz}, {Holz}, \&
  {Warren}}]{wetzl08}
{Wetzel}, A.~R., {Schulz}, A.~E., {Holz}, D.~E., \& {Warren}, M.~S. 2008, \apj,
  683, 1

\bibitem[{{White} {et~al.}(2005)}]{white05}
{White}, S.~D.~M. {et~al.} 2005, \aap, 444, 365

\bibitem[{{Williams} {et~al.}(2010){Williams}, {Quadri}, {Franx}, {van Dokkum},
  {Toft}, {Kriek}, \& {Labb{\'e}}}]{williams10}
{Williams}, R.~J., {Quadri}, R.~F., {Franx}, M., {van Dokkum}, P., {Toft}, S.,
  {Kriek}, M., \& {Labb{\'e}}, I. 2010, \apj, 713, 738

\bibitem[{{Willmer} {et~al.}(2006)}]{willmer06}
{Willmer}, C.~N.~A. {et~al.} 2006, \apj, 647, 853

\bibitem[{{Yan} \& {Blanton}(2011)}]{yan11b}
{Yan}, R. \& {Blanton}, M.~R. 2011, ArXiv e-prints

\bibitem[{{Yan} {et~al.}(2006){Yan}, {Newman}, {Faber}, {Konidaris}, {Koo}, \&
  {Davis}}]{yan06}
{Yan}, R., {Newman}, J.~A., {Faber}, S.~M., {Konidaris}, N., {Koo}, D., \&
  {Davis}, M. 2006, \apj, 648, 281

\bibitem[{{Yan} {et~al.}(2011)}]{yan11}
{Yan}, R. {et~al.} 2011, \apj, 728, 38

\bibitem[{{York} {et~al.}(2000)}]{york00}
{York}, D.~G. {et~al.} 2000, \aj, 120, 1579

\bibitem[{{Zhao}(2002)}]{zhao02}
{Zhao}, H. 2002, \mnras, 336, 159

\bibitem[{{Zirm} {et~al.}(2007)}]{zirm07}
{Zirm}, A.~W. {et~al.} 2007, \apj, 656, 66

\end{thebibliography}

\end{document}